\documentstyle[12pt,preprint,aps,epsfig]{revtex}
\newcommand{\npi}{\mbox{$\pi N$}}
\newcommand{\nsig}{\mbox{$\sigma N$}}
\newcommand{\nrho}{\mbox{$\rho N$}}
\newcommand{\neta}{\mbox{$\eta N$}}
\newcommand{\dpi}{\mbox{$\pi \Delta$}}

\newcommand{\sla}{ {\hspace{-0.19cm} \setminus} }
\newcommand{\slas}{ {\hspace{-0.19cm} \setminus}\hspace{-0.0cm} }
\begin{document}

\preprint{FZJ-IKP(TH)-99-30, NT@UW-99-58, and DOE/ER/40561-77-INT99}

\title{What is the structure of the Roper resonance?
}

\author{O. Krehl$^a$, C. Hanhart$^{b}$, S. Krewald$^a$, and J. Speth$^a$}

\address{$^a$ Institut f\"ur Kernphysik, \\
Forschungszentrum J\"ulich GmbH, 52425 J\"ulich,
Germany\\
$^b$ Department of Physics and INT,\\
University of Washington, Seattle, WA 98195, USA}

\date{\today}

\maketitle

\begin{abstract}
  We investigate the structure of the nucleon resonance $N^*(1440)$
  (Roper) within a coupled-channel meson exchange model for pion-nucleon
  scattering.  The coupling to $\pi\pi N$ states is realized
  effectively by the coupling to the $\sigma N$, $\pi \Delta$ and
  $\rho N$ channels. The interaction within and between these channels is derived
  from an effective Lagrangian based on a chirally symmetric
  Lagrangian, which is supplemented by well known
  terms for the coupling of the $\Delta$ isobar, the $\omega$ 
  meson and the '$\sigma$', which is the name given here to the strong
  correlation of two pions in the scalar-isoscalar channel. In this
  model the Roper resonance can be described by meson-baryon dynamics
  alone; no genuine $N^*(1440)$ (3 quark) resonance is needed in order to
  fit  $\pi N$ phase shifts and inelasticities. 

\end{abstract}

\section{Introduction}

The experimental and theoretical investigation of the baryon spectrum
helps to improve our knowledge of QCD in the nonperturbative regime --
especially of the confining mechanism, which is most important for
binding a system of quarks into a hadron.  Experimental information
about the mass, width and decay of baryon resonances serves as a testing
ground for several models of the internal structure of the nucleon and
its excited states. Most of this information is extracted from partial
wave analyses of $\pi N$ scattering data \cite{Koch85,SM95,SP98},
sometimes in combination with transition amplitudes to inelastic
channels such as $\pi N \to \eta N$\cite{Batinic95,Feuster99,Vrana99} or 
$\pi N \to
\pi\pi N$ \cite{Cutkosky79,Manley92,Vrana99}. In addition there is
information available form photo- and electro-production of $N^*$
resonances \cite{PDG98} or, as recently proposed, from the $N\bar{N}$ decay
channel
of the $J/\Psi$ \cite{Zou99}.

The mass spectrum of excited baryon states has been calculated within
several quark models (QM).  The nonrelativistic QM of Isgur and Karl
\cite{Isgur78}, for example,  leads to a good qualitative understanding of 
the negative parity resonances by assuming a structure of three
constituent quarks that are confined by a harmonic oscillator
potential and interact through  a residual interaction inspired
by one gluon exchange. In order to describe the positive parity states, however
they had to introduce an additional anharmonicity into the confining oscillator
potential that lowers the mass of the first positive parity
resonance ($N^*(1440)$) \cite{Isgur79}.  The relativized QM
\cite{Capstick86} gives a good qualitative picture of the baryonic
spectrum by using an interaction which, in the nonrelativistic limit, 
can be decomposed into a color Coulomb part, a confining
interaction, a hyperfine interaction and a spin-orbit interaction
between quarks. The confinement is provided by a Y-type string
interaction between all three quarks. One (of several) difficulties
with this model is that the low lying positive parity resonances are systematically overestimated by at least 100 MeV. A rather different 
interaction mechanism was used by Glozman and Riska \cite{Glozman96}. In 
their model, two quarks interact via pion exchange.
This flavor-dependent force is responsible for the low mass of the
Roper resonance ($N^*(1440)$). Confinement is achieved by an
oscillator potential.  Thus the interaction mechanisms of the
Glozman/Riska model and the Isgur/Karl/Capstick model are quite
different and it is not clear whether the mass spectrum should be
described by either one of these interactions or a mixture of both
\cite{Bijker97,Liu99,Lee99,Isgur99,Glozman99}.

The photo- and electro-excitation of baryon resonances have been studied by
several groups using several different models. Li and collaborators
\cite{Li90,Li92} found the $Q^2$ dependence of the $N^*\to N\gamma$
helicity amplitudes to be very sensitive to the structure of the Roper
resonance. While the nonrelativistic $q^3$ model is not able to
describe the $Q^2$ behavior, a hybrid $q^3g$ model is in agreement
with the available experimental data. A similar conclusion was reached
by Capstick \cite{Capstick92}, who found large disagreement in the
photo-production amplitude of the Roper between a theoretical
calculation in a nonrelativistic $q^3$ model -- including relativistic
corrections -- and the experimental data.  However Capstick and Keister
\cite{Capstick95} pointed out that relativistic effects are very
important in these amplitudes. They were able to describe the helicity
amplitudes using a ``relativized'' $q^3$ QM. Cardarelli et al. also
investigated the electro-production of the Roper resonance and
concluded that this resonance can hardly be interpreted as a simple
radial excitation of the nucleon \cite{Cardarelli97}.  Recently the
T\"ubingen group \cite{Dong99} found large contributions from
meson-baryon intermediate states in the transition amplitudes
$N^*(1440) \to N\gamma$. Thus even the study of its electromagnetic
excitation does not clearly reveal the structure of the Roper resonance.

The decay widths of baryons have been calculated using several approaches
by combining a QM with a model for the decay of the three quark system
into a meson baryon state, such as the $^3P_0$
model\cite{Capstick93,Capstick94}, or the string breaking mechanism of
the flux tube model \cite{Stancu89,Stancu90,Stassart92}. The $\pi N$ decay
width of the Roper resonance as calculated by Capstick and Roberts
\cite{Capstick93} is in agreement with the analysis of Cutkosky and Wang
\cite{Cutkosky90} but, compared to the partial wave analysis of the
Karlsruhe \cite{Koch85} and the VPI \cite{SM95,SP98} groups, the decay
width of the Roper should be much smaller. In addition, none of the
decay models include any kind of meson-baryon final state interaction
or coupled-channel effects \cite{Capstick93}, although there are
indications that these could lead to large shifts of the energy levels
and mixing effects between states \cite{Dong99,Barnes94}. A
consistent investigation of higher Fock states, such as $q^4\bar{q}$, is
missing \cite{Capstick86}, although there are investigations of
$q^4\bar{Q}$ systems, where $Q=s$ \cite{Maltman86} or $Q=c,b$
\cite{Stancu98,Genovese98}.

At this stage a closer look at the different partial wave analyses may
help us to understand the problem in more detail. In
Table~\ref{tabp11} we have listed the mass, width and pole position of
the Roper resonance as extracted from several partial wave analyses of
$\pi N$ scattering data.  The first five lines correspond to models
that either get the mass, $m_R$, and width, $\Gamma$, of the Roper
resonance by fitting a Breit-Wigner-like resonance to the $\pi N$ data
or derive the position of the resonance pole in the complex
energy plane. This pole position can be related to the mass and width
of the resonance by
\begin{equation}\label{poleq}
m_R=Re(\mbox{Pole}), \quad \Gamma=-2Im(\mbox{Pole}),
\end{equation}
which, in fact, is the origin of the denominator in a Breit-Wigner
parameterization of a resonance.  By comparing the mass and width
parameters of the analyses a) -- e) to the position of the pole as
found in a), b), d), and e) one can see large discrepancies. The mass,
as extracted from the pole, lies typically $\approx$ 100 MeV below
$m_R$. Something similar can be seen by comparing the widths: here a
ratio $-\frac{\Gamma}{Im(\mbox{Pole})} \approx 5$ is found instead of
the expected value of 2. For an undistorted resonance, such as the
$N_{D_{13}}^*(1520)$, the mass and width from the Breit-Wigner
parameterization and the pole position are essentially the same within a few
MeV \cite{PDG98}. This observation shows already that the Roper
resonance is substantially influenced by strong meson-baryon
background interactions and/or effects from nearby thresholds.
H\"ohler suggested the use of the pole position as source of information
on the mass and width of a resonance, since the pole has a
well-defined meaning in $S$-matrix theory \cite{Hoehler98}.  If we do
so, the QMs use the wrong values for the mass and width of the Roper
resonance. Compared to the pole position values of $m_R$ and $\Gamma$
(calculated using Eq. (\ref{poleq})), the relativized QM
\cite{Capstick86} overestimates the mass of the Roper by about 200
MeV and the $\pi N$ decay width of the Roper resonance is
overpredicted too.

\begin{figure}
\begin{center}
\epsfig{file=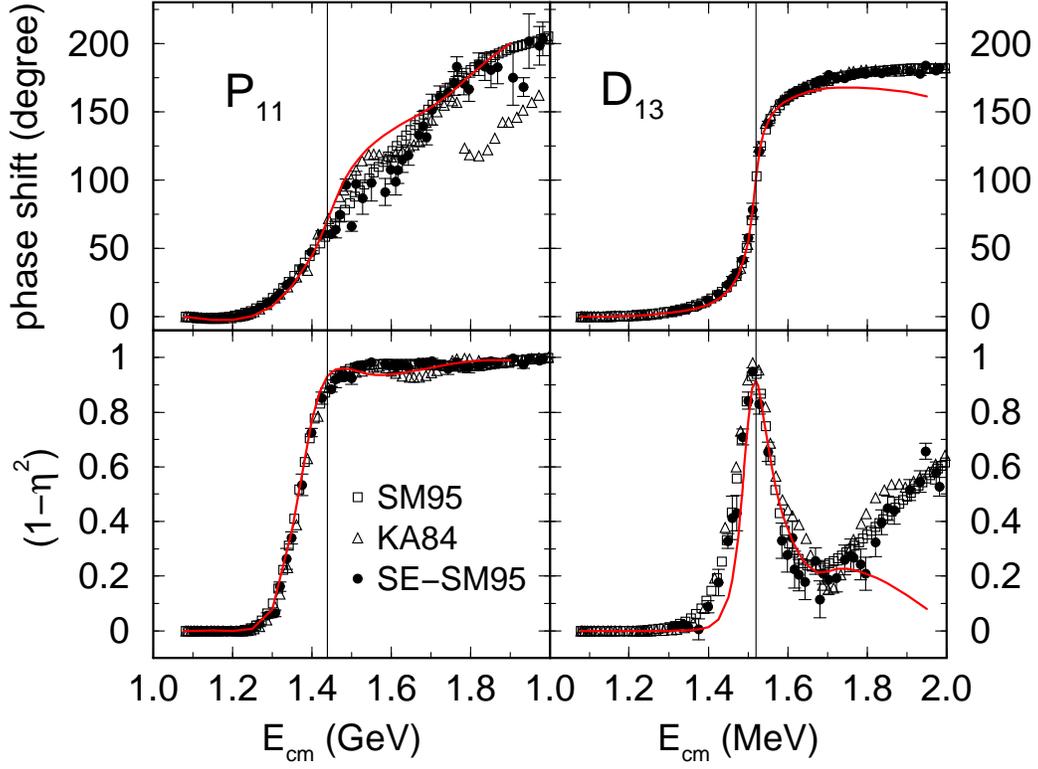,height=10.5cm}
\caption{Phase shift and inelasticity in the partial waves $P_{11}$ and
$D_{13}$. Data are
  taken from Ref. \protect \cite{SM95} (SM95) and \protect
  \cite{KA84,Koch85} (KA84). In addition, the single-energy analysis
  from \protect \cite{SM95} (SE-SM95) is shown. The vertical lines are
  drawn at $E=1440$ MeV ($P_{11}$) and $E=1520$ MeV ($D_{13}$) and
  correspond to the suggested values of the resonance masses as given in Ref.
  \protect \cite{PDG98}.}
\label{phasep11}
\end{center}
\end{figure}

Another remarkable difference between the $N^*(1520)$ and the
$N^*(1440)$ is seen in examination of the partial wave amplitudes
(displayed as phase shift $\delta$ and inelasticity $\eta$) in Fig.
\ref{phasep11}. The $N^*(1520)$ causes a nice change in the phase
shift of the partial wave $D_{13}$ up to $180^o$ and crosses $90^o$ at
$\approx$ 1520 MeV. This is also the position of the maximum in the
inelasticity. After passing the resonant phase of $90^o$,
the amplitude goes back to being almost elastic. The
situation is completely different for the $N^*(1440)$. Here the phase
shift in the $P_{11}$ increases slowly, which corresponds to a very
broad resonance, but the inelasticity opens very rapidly (almost as
fast as in the $D_{13}$) and remains inelastic over a very large energy
range. Furthermore, the suggested resonance position of $m_R=1440$ MeV
does not correspond to $\delta=90^o$. The shape of the
$P_{11}$ partial wave amplitude in the region of the Roper resonance
also looks very different from a typical Breit-Wigner resonance. To 
summarize, the Roper appears not to fit into our picture of
Breit-Wigner-like resonances.

A series of different methods can be found in the literature that try
to extract information on the Roper resonance from $\pi N$ scattering.
The ones displayed in Table \ref{tabp11} can be summarized as follows: 
\begin{itemize}
\item Analyses a) and b) are combined analyses of all available $\pi N$
  scattering data.  Two methods are used in order to extract
  parameters of resonances.  First, a coupled-channel $K$-matrix approach, additionally
  constrained by fixed $t$ dispersion relations, allows
  a continuation of the partial wave amplitudes into the complex energy
  plane, where the poles of the resonances can be found. Second,
  fits to single-energy partial wave solutions using generalized
  Breit-Wigner parameterizations are performed, which lead to the
  values of $m_R$ and $\Gamma$.
\item Manley and Saleski (c) use a combination of Breit-Wigner
  resonances and a phenomenological parameterization of the
  background, which is unitarized in a $K$-matrix approximation. They
  included experimental data of the reaction $\pi N \to \pi\pi N$ into
  their fitting procedure.
\item The group of Cutkosky (d) used a separable coupled-channels 
  resonance model.  The dressed propagator of the intermediate
  resonances is a solution of the Dyson equation and the vertices are
  generalized Breit-Wigner vertex functions. Backgrounds are
  parameterized as resonance contributions with a resonance position
  below threshold.
\item Analysis (e) is an extended version of the model used in (d).
  Input data are the partial wave solutions of the VPI group
  \cite{SM95} and the transition cross sections $\pi N \to \eta N$ and
  $\pi N \to \pi\pi N$.
\item In f), g), and h) H\"ohler and Schulte use the speed plot
  method for determining resonance parameters. We describe this
  method in more detail in Sec. \ref{secroper}. The speed plot analysis uses
other
  partial wave solutions as input and therefore is not a partial wave analysis
of
  $\pi N$ scattering, but an alternative way of extracting resonance parameters.
\item Line (i) represents our results, which will be discussed in
  detail in Sec. \ref{secroper}.
\end{itemize}
All of these analyses agree in the need for a pole in the partial wave
$P_{11}$ and all of them but our work assume a small background interaction.
However the aim of analysis a) -- h) is {\it not} to determine the structure of a
resonance. This was pointed out in a recent extension of the CMB model
by Vrana, Dytman, and Lee \cite{Vrana99}. Rather, these analyses seek
to discover whether there is a resonance or not.  They do so by
providing the poles demanded by data as input.  The number of poles as
well as their parameters are then obtained by means of a $\chi^2$ fit.

In addition to these analyses there are many theoretical models for $\pi N$
scattering up
to the energies of the first $N^*$ resonances. They can be divided into two
classes:
\begin{description}
\item[Separable potential models] such as \cite{Blankleider85,Fuda98}.
  In these models the potential $V$ of a coupled-channel
  Lippmann-Schwinger equation (LSE) is assumed to be of the separable
  form $V(k',k)=f(k')\lambda f(k)$, where k (k') is the relative
  momentum of the initial (final) state. The form factor $f$ is
  parameterized differently for each partial wave, and the strength
  factor $\lambda$, together with the  parameters of the form factor, 
  is adjusted to fit data. Since the parameters of the
  form factors do not have a clear physical meaning, the interpretation
  of these parameters in terms of resonances and backgrounds is not
  possible. Nevertheless, one can still learn about effects of opening
  thresholds of coupled-channels.

\item[$K$-matrix approximations,] such as the models introduced in Refs.
  \cite{Feuster99,Gridnev99}. These use a microscopic potential, $V$, as
  input to a LSE, which is solved in the
  $K$-matrix approximation. In general, a LSE (written in a symbolic notation)
\begin{equation}\label{lse}
T=V+V\frac{1}{E-H^0+i\epsilon}T
\end{equation}
can be decomposed into a set of equations
\begin{eqnarray}
K&=&V+V\frac{{\cal P}}{E-H^0}K,\label{kgl}\\
T&=&K-i\pi K \delta(E-H^0)T,\label{tgl}
\end{eqnarray}
where we have introduced the $K$-matrix \cite{Joachain75,Machleidt87}
and ${\cal P}$ denotes the principal value. The $K$-matrix
approximation now simplifies this set of equations by setting $K=V$.
This reduces the integral equation (\ref{lse}) to an algebraic
equations (\ref{tgl}). The $K$-matrix approximation does not allow for
virtual intermediate states. One consequence of this is, that the
different channels only contribute above their production threshold.
Of course this truncates the strength of the virtual states and,
consequentially, the strength of the multiple scattering
contributions. This can also be found in a slightly more formal way:
The Heitler equation, Eq. (\ref{tgl}), introduces the unitary cut to
the $K$-matrix so that the $T$-matrix contains this unitary cut and
the poles present in $K$.  The rescattering of virtual states is
described completely by the $K$-matrix Eq. (\ref{kgl}). Since this is
a Fredholm type of integral equation, it can be solved by iteration
\begin{equation}\label{reihe}
K=V+V\frac{{\cal P}}{E-H^0}V+V\frac{{\cal P}}{E-H^0}V\frac{{\cal
P}}{E-H^0}V+\dots
\end{equation}
This series may be divergent \footnote{It is when there is a bound
  state at the energy at which this equation is solved.}, which
introduces (besides the poles in $V$) additional poles due to
rescattering.  These poles are not present if the $K$-matrix equation
is approximated by cutting off the series (\ref{reihe}) at a finite
order. Even if no pole is generated by the infinite sum, there may
still be much strength in higher order iterations, which are
eliminated in approximating $K=V$.  With this in mind, it is clear,
that the $K$-matrix approximations discussed above do not find
dynamical poles such as bound states.
\end{description}

It has long been known that the poles of the two-body $S$-matrix (as a
function of a complex energy variable) are not only resonance poles,
but can also be bound state poles or coupled-channel poles
\cite{Badalyan82}. A bound state is generated by a strongly attractive
interaction between two particles, whereas a coupled-channel pole can
be realized by a coupling between two reaction channels. Prominent
examples of bound states of two hadrons are the $f_0(980)$, which is
found to be a $K\bar{K}$ molecule in the $\pi\pi / K\bar{K}$ system
\cite{Weinstein90,Janssen95} and the $\Lambda(1405)$ as $\bar{K} N$
bound state in the $\pi \Sigma/\bar{K} N$ system
\cite{Siegel88,Groeling90}. An example of a coupled-channel pole can
be found in the $\pi\eta / K\bar{K}$ system, where the $a_0(980)$ can
be generated by the coupling between these two channels
\cite{Janssen95}. It is, however, not always easy to distinguish
between these two types of poles.

The situation we have presented so far can be summarized as follows:
The QM calculations do not give us a clear picture of the structure
of the Roper resonance, even by studying electromagnetic processes
or decay widths. Yet we know that in many analyses of $\pi N$
scattering the need of a resonance has been found. The aim of
these analyses was not to determine the structure of the resonance,
but to determine resonance parameters, such as masses, widths and
branching ratios.  The coupled-channel models of $\pi N$ scattering
for energies under consideration work in the $K$-matrix
approximation, in which part of the strength due to virtual intermediate
states is truncated. Furthermore the $\pi\pi N$ states in these
models are not treated consistently; rather, the mass of some effective
$\pi\pi N$ channel is adjusted differently in each partial wave
\cite{Gridnev99}, or an unphysical scalar-isovector $\pi\pi$ state is
used \cite{Feuster99}.

A model for $\pi N$ partial wave amplitudes as solution of a full LSE
up to energies of 1.9 GeV is missing. Our aim is therefore to
construct such a model in order to investigate whether or not it is
possible to describe the Roper resonance as a dynamically generated
resonance. We use the model of Ref. \cite{Schuetz98} as a
starting point. This model is able to describe the $\pi N$ partial
waves up to energies of 1.6 GeV by coupling the channels $\pi N,\sigma
N,\pi \Delta$, and $\eta N$ and has proven its ability to analyze
the structure of a resonance in the partial wave $S_{11}$ and
$P_{11}$.  We have improved this model in several significant ways:
\begin{itemize}
\item We have included the $\rho N$ reaction channel into the
  coupled-channel calculation in order to complete the effective
  description of $\pi\pi N$ states. This channel improves the
  description in the partial waves $P_{13}$ and $P_{31}$ and leads to
  large contributions in the partial wave $S_{11}$ in the region of
  the $N^*(1650)$.
\item In Ref. \cite{Schuetz98} $t$ channel $\pi$ exchange diagrams
  were omitted in order to avoid double counting. By dropping these
  terms also the coupling strength between the $\pi N$ and the
  $\sigma N$ channel is weakened. We have included
  these diagrams (Fig. \ref{ropgraf} (j) and \ref{rhongraf} (a))
  explicitly and avoid the double counting problem by modifying the
  $N\bar{N} \to \pi\pi$ amplitudes (see Sec.  \ref{secmodel} for more
  details). This results in a large coupling between the
  $\pi N$ and $\sigma N(\rho N)$ channels, which was not present in
  \cite{Schuetz98}. 
\item The rules of time ordered perturbation theory were applied with care,
  which leads to additional contact interactions (see the appendix
  for more details). In \cite{Bockmann99} these contact terms
  are found to be large corrections and we also find strong
  contributions of these additional interactions e.g. in the $\pi $
  exchange diagrams.
\end{itemize}

In the next section our model is described in greater detail.
In Sec. \ref{secdata} we shall discuss the results of this model as
compared to the amplitudes of partial wave analyses and some
transition cross sections.  Sec. \ref{secroper} will be dedicated
to an investigation of the structure of the Roper resonance. The
last section summarizes our results.

\section{$\pi N$ scattering in a meson exchange model
}\label{secmodel}

In the introduction we argued that a detailed investigation of
the Roper resonance goes along with an understanding of $\pi N$
scattering over a rather large energy region -- from threshold
($E=\sqrt{s}=1077$ MeV) up to energies well above the resonance under
investigation (e.g., 1.9 GeV). Furthermore we have to use a realistic
interaction between the meson and the baryon. Such an interaction is
provided by the meson exchange model, which has successfully been used
in many different reactions such as the $NN$ interaction \cite{Machleidt87},
the elastic $\pi N$ interaction,
\cite{Pearce91,Lee91,Gross93,Hung94,Goudsmit94,Pascalutsa98,Lahiff99},
the $KN$ interaction \cite{Buettgen90}, the $\bar{K}N$ interaction
\cite{Groeling90} and the $\pi\pi$ interaction \cite{Janssen95}, to name just
a few. Before we go into the details of the interaction, we
wish to specify the reaction channels we will need in our description.

From Fig. \ref{phasep11} it is clear that the $\pi N$ interaction
above energies of 1.3 GeV is very inelastic.  The decay modes of the
nucleon resonances in the energy range under consideration show that
the dominant decay (besides $\pi N$ and $\eta N$ for the $N^*(1535)$)
is the $\pi\pi N$ channel \cite{PDG98}. Since a three-body calculation is much
too complicated for realistic potentials, we must reduce the $\pi\pi N$
channel into effective two-body channels. In doing this we are guided by
studying strong interactions between two-body clusters of the three-body $\pi\pi
N$ state. The dominant clusters are the $\Delta$ in the $\pi N$
interaction, the $\rho$ in the vector isovector $\pi\pi$ interaction and the
strong correlation in the scalar-isoscalar $\pi\pi$ interaction, which we
call $\sigma$. Therefore -- besides the $\pi N$ and $\eta N$ channels, which
are needed for a
complete description of the $N^*(1535)(S_{11})$ -- our model includes
the reaction channels $\pi\Delta$, $\sigma N$ and $\rho N$.

We have then to solve the coupled-channel scattering
equation \cite{Groeling90}
\begin{eqnarray}\label{ccstreugl}
T^I_{\mu\nu}(\vec{k}',\lambda_3,\lambda_4;\vec{k},\lambda_1,\lambda_2)&=&V^I_{\mu\nu}(\vec{k}',\lambda_3,\lambda_4;\vec{k},\lambda_1,\lambda_2)+
 \nonumber \\
&&\hspace{-2cm}\sum_{\gamma}\sum_{\lambda'_1,\lambda'_2}\int d^3q
V^I_{\mu\gamma}(\vec{k}',\lambda_3,\lambda_4;\vec{q},\lambda'_1,\lambda'_2)
\frac{1}{E-W_{\gamma}(q)+i\epsilon}T^I_{\gamma\nu}(\vec{q},\lambda'_1,\lambda'_2;\vec{k},\lambda_1,\lambda_2),
\end{eqnarray}
where $\lambda_i,\lambda_{i+2},\lambda'_i, (i=1,2)$ are the helicities of the
baryon and meson in the initial, final and intermediate state, $I$ is
the total isospin of the two body system and $\mu,\nu,\gamma$ are indices 
that label different reaction channels.
$W_{\gamma}(q)=\sqrt{q^2+M_{\gamma}}+\sqrt{q^2+m_{\gamma}}$ where
$m_{\gamma}(M_{\gamma})$ is the mass of the meson (baryon) in the
channel $\gamma$, respectively. We work in the center-of-momentum (cm)
frame and
$k(k')$ are the momenta of the initial (final) baryon, respectively.

The pseudopotential $V$ (i.e., the interaction between baryon and
meson) that is iterated in Eq. (\ref{ccstreugl}) can be constructed
from an effective Lagrangian. Our interaction Lagrangian (see Table
\ref{tablag}) is based on that of Wess and Zumino \cite{Wess67},
which we have supplemented with additional terms for including the
$\Delta$ isobar, the $\omega$, $\eta$, $a_0$, $f_0$ meson and the $\sigma$.
We also have included terms that characterize the coupling of
the resonances $N^*(1535)$, $N^*(1520)$ and $N^*(1650)$ to various
reaction channels. The full interaction is built up by the diagrams
shown in Figs. \ref{pingraf}--\ref{rhongraf}, where we also
introduce our notation.  Expressions for the matrix elements $\langle
\vec{k}' \lambda_3\lambda_4|V^{IJ}|\vec{k} \lambda_1\lambda_2\rangle$ can be
found in the Appendix.

In our approach the correlated $\pi\pi$ exchange replaces the exchange
of fixed-mass $\rho$ and $\sigma$ mesons. The construction of these
potentials is explained in detail in Ref. \cite{Krehl99}.  However
double counting will arise when correlated $\pi\pi$ exchange and the
$\pi$ exchange diagrams in the $\pi N \to \sigma (\rho) N$ transition
potential are taken into account \cite{Schuetz98}. For this reason
Sch\"utz et al. \cite{Schuetz98} left out the $\pi$ exchange
contributions. But these diagrams are important contributions to the
$\pi N\to \sigma(\rho) N$ potential and therefore have to be included
in our model. We avoid the double counting, which arises by iterating
the $\pi$ exchange diagrams (see Fig. \ref{fdouble}) by modifying the
$N\bar{N} \to \pi\pi$ amplitudes.  Since we have a microscopical model
for the $N\bar{N} \to \pi\pi$ $T$-matrix \cite{Schuetz95}, we are able
to subtract the box diagram displayed in Fig. \ref{fdouble} c) from
these amplitudes.  When using the subtracted amplitudes
$T_{\mbox{corr}}$, double counting is avoided. The subtraction of the
box diagram hardly influences the $\rho$ partial waves in the
$N\bar{N} \to \pi\pi$ amplitudes, whereas it reduces the $\sigma$
channel by $\approx$ 20 \%.  By solving the double counting problem in
this way we can keep the important $\pi$ exchange diagrams in the $\pi
N \to \sigma (\rho) N$ transition amplitudes.

After a standard partial wave decomposition \cite{Erkelenz74}, the
scattering equation (\ref{ccstreugl}) can be reduced to a one-dimensional 
integral equation that can be solved by standard methods
\cite{Hetherington65,Aaron66,Cahill71}. A unitary transformation
relates the helicity states we have used in Eq. (\ref{ccstreugl}) to the so
called $JLS$ states \cite{JacobWick59,Schuetz94}.
In the $JLS$ basis the $T$-matrix is directly
related to the partial wave amplitudes \cite{Hoehler83}
\begin{equation}\label{pwamp}
\tau_{\mu\nu}^{IJLSL'S'}=-\pi\sqrt{\rho_{\mu}\rho_{\nu}}T^{IJLSL'S'}_{\mu\nu}
\end{equation}
where the densities $\rho_{\gamma}$ are given by
$\rho_\gamma=\frac{q^{\gamma}_{on}}{E}E_{\gamma}(q^{\gamma}_{on})\omega_{\gamma}(q^{\gamma}_{on})$,
with
$E_{\gamma}(k)=\sqrt{k^2+M_{\gamma}^2}$, $\omega_{\gamma}=\sqrt{k^2+m_{\gamma}^2}$
and
$q^{\gamma}_{on}=\sqrt{[E^2-(M_{\gamma}+m_{\gamma})^2][E^2-(M_{\gamma}-m_{\gamma})^2]}/2E$.
Here $JLS$ are the usual total angular momentum, orbital angular momentum and total
spin quantum numbers and the prime denotes final state quantities.
For the partial wave amplitudes in which we are mostly interested in this work,
namely the $\pi N$ amplitudes, the total spin $S$ and orbital angular
momentum $L$ are conserved ($L'=L$, and $S'=S=1/2$ for $\mu=\nu=\pi N$) in Eq.
(\ref{pwamp}). The phase shift and inelasticity are then calculated from
the partial wave amplitude in the standard way \cite{Hoehler83}.

Mesons and baryons are not point-like particles, but have a finite size.
Therefore the interaction vertices $mmm$ and $mBB$ ($m$=meson,
$B$=Baryon) also have a finite sizes which, in our model, are parameterized by the
following form factors, in which $\vec{q}$ is the three momentum transfer
carried by the exchanged particle:
\begin{itemize}
\item For meson and baryon exchange
\begin{equation}\label{ffaus}
F(q)=\left(\frac{\Lambda^2-m_x^2}{\Lambda^2+\vec{q}\,^2}\right)^n .
\end{equation}
We use monopole form factors ($n=1$) except for the $\Delta$
exchange, for which the convergence of the integral in Eq. 
(\ref{ccstreugl}) requires a dipole form factor ($n=2$).
\item For the nucleon exchange at the $\pi NN$ vertex
\begin{equation}\label{ffnu}
F(q)=\frac{\Lambda^2-m_N^2}{\Lambda^2-((m_N^2-m_{\pi}^2)/m_N)^2+\vec{q}\,^2}.
\end{equation}
This choice ensures that the nucleon pole and nucleon exchange
contribution cancel each other at the Cheng-Dashen point, which is needed for a 
calculation of the $\Sigma$ term \cite{Schuetz94}.
\item For $N$, $N^*$ and $\Delta$ Pole diagrams
\begin{equation}\label{ffpol}
F(q)=\frac{\Lambda^4+m_R^4}{\Lambda^4+(E_{\gamma}(q)+\omega_{\gamma}(q))^4}.
\end{equation}
\item The correlated $\pi\pi$ exchange is supplemented by the form factor
\begin{equation}\label{ffcorr}
F(t,t')=\left(\frac{\Lambda^2-t'}{\Lambda^2-t}\right)^2,
\end{equation}
which appears inside the $t'$ integration \cite{Schuetz94}.
\item For the contact interaction in the Wess-Zumino Lagrangian \cite{Wess67}
\begin{equation}\label{ffct}
F(p_2,p_4)=\left(\frac{\Lambda^2+m_4^2}{\Lambda^2+\vec{p}_4\,^2}\frac{\Lambda^2+m_2^2}{\Lambda^2+\vec{p}_2\,^2}\right)^2.
\end{equation}
\end{itemize}

All of our effective $\pi\pi N$ states (i.e., $\pi \Delta$, $\sigma N$
and $\rho N$) are composed of a stable and an unstable particle. In
order to include effects of the width of these unstable intermediate states we
have modified the two-body propagator, which will be motivated in the
following. Since in the Schr\"odinger equation,
\begin{equation}\label{heq}
H|\Psi\rangle=E|\Psi\rangle,
\end{equation}
the Hamilton operator acts on Hilbert states describing a particle $R$ as well
as two particles $12$ into which $R\to12$ can decay, we introduce
Feshbach projectors
\begin{equation}
P=|R\rangle\langle R|, \quad
Q=|12\rangle\langle 12|, \quad \mbox{with}\quad
P+Q=1, \quad
P^2=P, \quad
Q^2=Q \quad
\end{equation}
in order to split these two spaces \cite{Afnan98,Elsey89}. By applying
these operators to the Eigenvalue equation (\ref{heq}), one can
derive an equation for the particles in $P$ space
\begin{equation}\label{peq}
\left(E-H_{PP}-H_{PQ}\frac{1}{E-H_{QQ}}H_{QP}\right)|\Psi_P\rangle=0,
\end{equation}
where $|\Psi_P\rangle=P|\Psi\rangle$ and $H_{XY}=XHY$. By introducing
the self-energy
\begin{equation}
\Sigma=H_{PQ}\frac{1}{E-H_{QQ}}H_{QP}
\end{equation}
equation (\ref{peq}) can be rewritten as
\begin{equation}
(E-H^0-\Sigma)|\Psi_P\rangle=0
\end{equation}
The self-energy term takes the decay of the unstable particle into
account. As such it introduces an energy-dependent width and a mass
shift. Our two-particle intermediate state propagator for $\pi\Delta$,
$\sigma N$ and $\rho N$ must therefore be replaced by
\begin{equation}\label{modprop}
\frac{1}{E-W_{\gamma}(q)}\to\frac{1}{E-W_{\gamma}(q)-\Sigma_{\gamma}(E_{sub})},
\end{equation}
where
\begin{eqnarray}
E_{sub}&=&E-\omega_{\pi}(q)-(\sqrt{(M_{\Delta}^o)^2+q^2}-M_{\Delta}^o) \mbox{
for the } \Delta, \nonumber \\
E_{sub}&=&E-E_{N}(q)-(\sqrt{(m_r^o)^2+p^2}-m_r^o)  \mbox{ for } r=\rho,\sigma
\end{eqnarray}
is the energy of the decaying cluster at rest \cite{Schuetz98}.  After
constructing models for the self-energies $\Sigma$, the bare masses
$M^o_{\Delta}$ and $m_r^o$ (as free parameters within these models)
are determined by fitting the models to experimental data.  For
simplicity we use separable interactions for calculating the
self-energy. For the $\Delta$ and the $\sigma$ this has already been
done in Ref. \cite{Schuetz98}, from which we take the self-energies
$\Sigma_{\gamma} (\gamma=\Delta,\sigma)$. For the $\rho$ we use the
vertex function
\begin{equation}
v^0_{\rho\pi\pi}(q)=\frac{g_{\rho\pi\pi}}{2\pi\sqrt{3}}
\frac{q}{\omega_{\pi}(q)\sqrt{\omega_{m_\rho^0}(q)}}
\frac{\Lambda_{\rho}^2+m_{\rho}^2}{\Lambda_{\rho}^2+4(\omega_{\pi}(q))^2}
\end{equation}
with the parameters
\begin{equation}
\frac{g^2_{\rho\pi\pi}}{4\pi}=2.9,\quad \Lambda_{\rho}=1.8 \mbox{GeV},
\quad m_{\rho}^0=911 \mbox{MeV}.
\end{equation}
With this vertex function the self-energy $\Sigma_{\rho}$ can be
calculated in the same way as outlined for the $\sigma$ in Ref.
\cite{Schuetz98} (see also Eq. (\ref{self}) below).  Fig.
\ref{selfrho} shows our separable interaction for the $\rho$ decay compared
with $\pi\pi$ scattering data.

This completes our model. The $\pi N$ partial wave amplitudes are
calculated by solving the LSE (\ref{ccstreugl}) with the propagator
(\ref{modprop}) for unstable intermediate states. The pseudopotential
$V$ is derived from the Lagrangian of Table \ref{tablag}.  Its parameters 
are the coupling constants and cutoffs for each
vertex that we have listed in Table \ref{tabparams}.

\section{Description of $\pi N$ data
}\label{secdata}
\begin{figure}[h]
\begin{center}
\epsfig{file=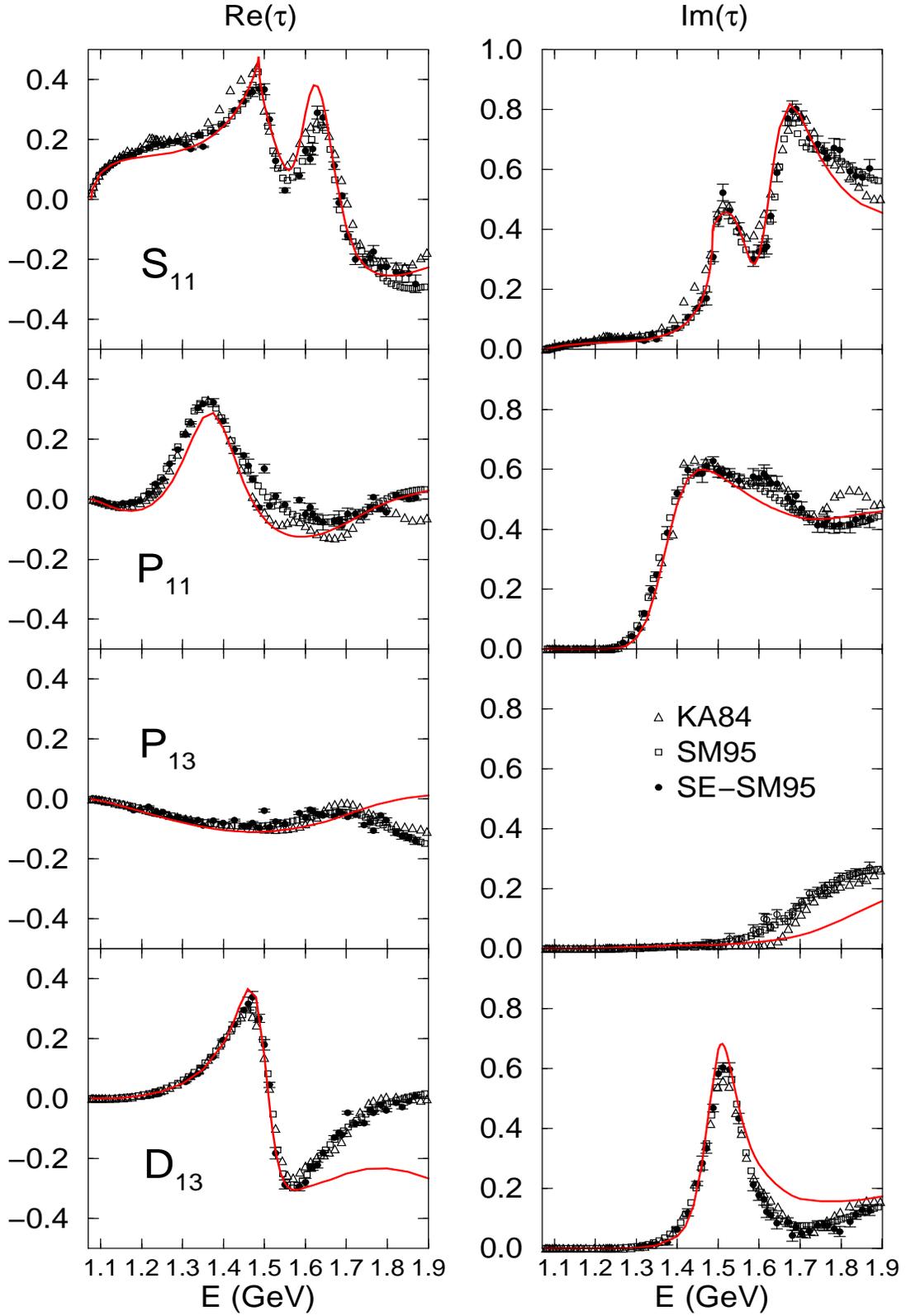,height=21.5cm,width=14.5cm}
\caption{The $\pi N$ partial wave amplitudes for the isospin $I=\frac{1}{2}$.
  In addition, the analyses KA84 \protect
  \cite{KA84,Koch85} and SM95 \protect \cite{SM95}, as well as the single-energy
analysis SE-SM95
\protect \cite{SM95} are shown.}
\label{pwa1}
\end{center}
\end{figure}

\begin{figure}[h]
\begin{center}
\epsfig{file=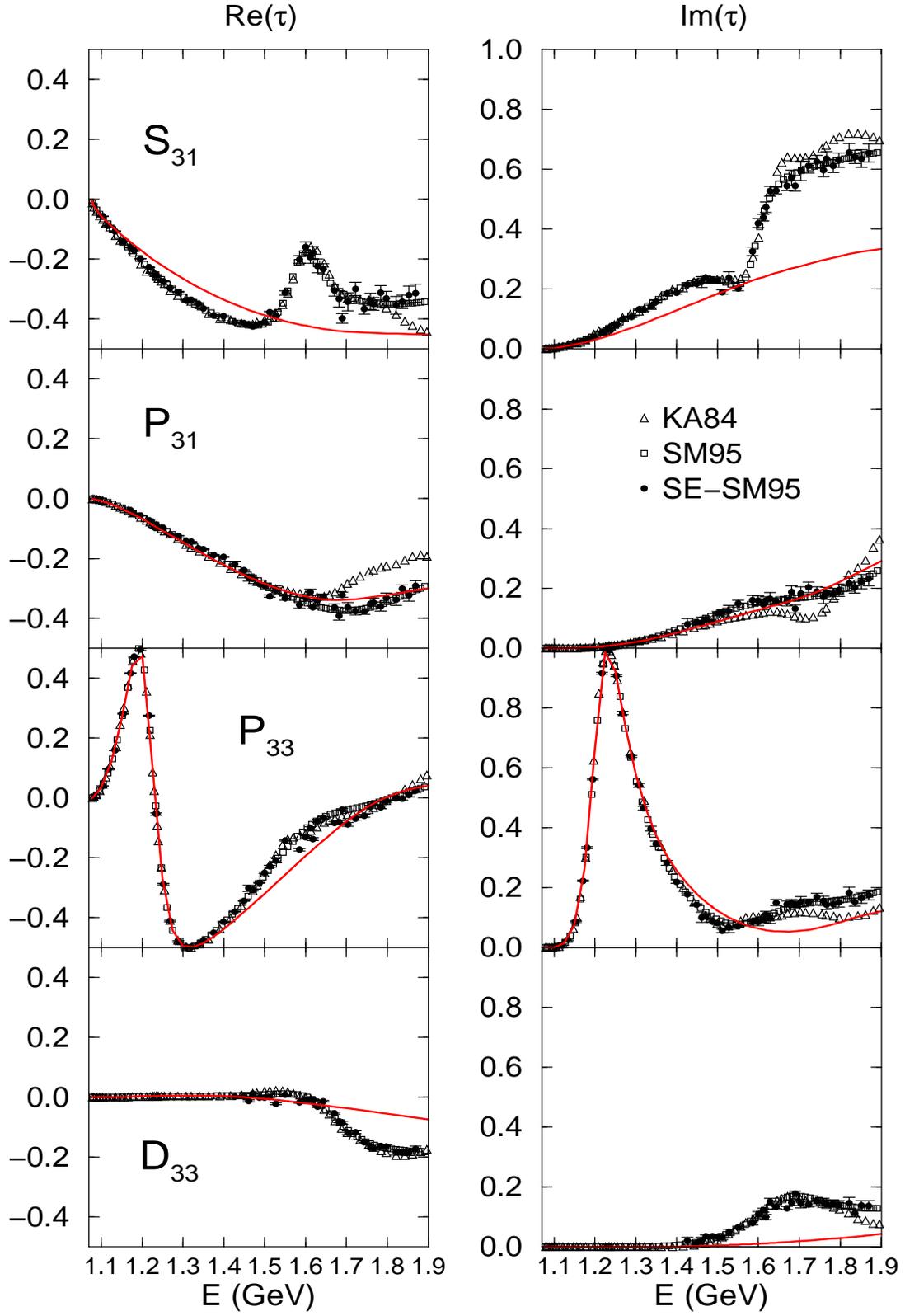,height=21.5cm,width=14.5cm}
\caption{The partial wave amplitudes for $I=\frac{3}{2}$.
The notation is the same as in Fig. \ref{pwa1}.}
\label{pwa3}
\end{center}
\end{figure}

Having described our model, we turn now to comparing its results to 
the experimental data. In fitting the partial wave amplitudes for
$J<\frac{5}{2}$ we have varied only the boldface printed values in
Table  \ref{tabparams}. Most of the coupling constants have been taken
from other sources. The coupling constants of the pole diagrams are
constrained by values determined from their decay widths, for which we
take the estimates of Ref \cite{PDG98}. The free values are then
strongly constrained by the data -- especially for the nonresonant $t$
and $u$ channel contributions, which act simultaneously in many
partial waves.  For completeness, Table  \ref{tabmassen} contains the
masses of the particles used in this model.  Our description of the
partial waves with $I=\frac{1}{2}$ is shown in Fig.  \ref{pwa1}; the 
partial wave amplitudes for $I=\frac{3}{2}$ are shown in  Fig.  \ref{pwa3}.

\begin{figure}[h]
\begin{center}
\epsfig{file=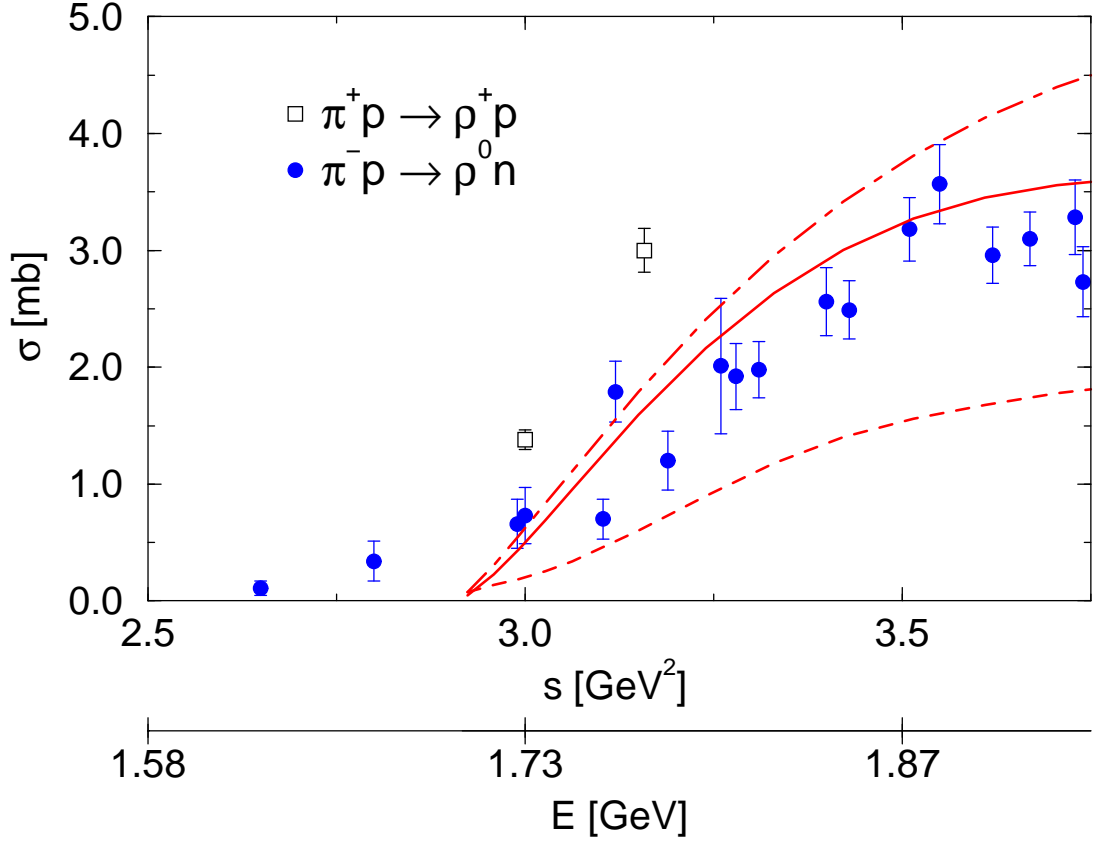,height=11.5cm}
\caption{The transition cross section $\pi N \to \rho N$. The solid line shows
the reaction
$\pi^- p \to \rho^0 n$, the dashed line the reaction
$\pi^- p \to \rho^- p$ and the  dot-dashed line the reaction
$\pi^+ p \to \rho^+ p$.
The experimental data are taken from Ref. \protect \cite{LB87}.}
\label{pirhocr}
\end{center}
\end{figure}

In order to constrain the parameters of the $\pi N \to \rho N$
transition potential, we have also considered the $\pi N \to \rho N$
transition cross section (Fig. \ref{pirhocr}). These data
severely constrain the $\pi$ exchange (Fig. \ref{rhongraf} a), which
dominates this cross section and produces a large background to the
resonant part in the $D_{13}$.  Without constraining the $\pi$
exchange contribution, a dynamical pole can be generated in the
$D_{13}$. This result was also obtained by Aaron et al.
\cite{Aaron68,Aaron69}. With this dynamical pole our model
overestimates the $\pi N \to \rho N$ cross section by almost an order
of magnitude, and a good description of other $\pi N$ partial waves is
not possible.  This demonstrates that only a combined analysis of many
partial waves and cross sections can give reliable information about
resonances. The details of this calculation will be presented
elsewhere \cite{Krehl2000}.

Our model is able to describe $\pi N$ data very well up to energies of
about 1.9 GeV.  Only in the partial wave $S_{31}$ does our model deviate
from the data, and that is because we have not yet included the resonance
$\Delta(1620)$. Our model does not give significant contributions to
the inelasticity in this partial wave.  The description of the
$S_{11}$ needs the coupling to the $\eta N$ channels via the
$N^*(1535)$ resonance and nonresonant $a_0(980)$ exchange
\cite{Schuetz98,Gridnev99}. The resonance $N^*(1650)$ is taken into
account in addition and leads to the rapid variation of the partial
wave amplitude around 1.65 GeV.  The inclusion of the $\rho N$
channel improves the description of the partial waves $P_{13}$ and
$P_{31}$ as compared to the model used in Ref. \cite{Schuetz98}, which
results in a perfect description of the $P_{31}$, whereas in the
$P_{13}$ a large background to the resonance $N^*(1720)$ is produced.
These results will be discussed in more detail elsewhere \cite{Krehl2000}.

The model is then a good starting point for an investigation of the Roper
resonance.

\section{The structure of the Roper resonance}\label{secroper}

Let us begin this section with a description of our procedure for 
investigating the structure of a resonance. We start by
using nonresonant interactions only; i.e., we do not include a pole
diagram into our interaction.  If we are able to fit data in all
partial waves without pole diagram, the resonance under consideration
does {\it not} have a three-valence-quark structure. Rather, it is created
dynamically
by the nonresonant meson-baryon interaction. If we need to include a
pole diagram, we conclude that the resonance is dominated by quark 
gluon dynamics, which are not included in our model.

\begin{figure}[h]
\begin{center}
\epsfig{file=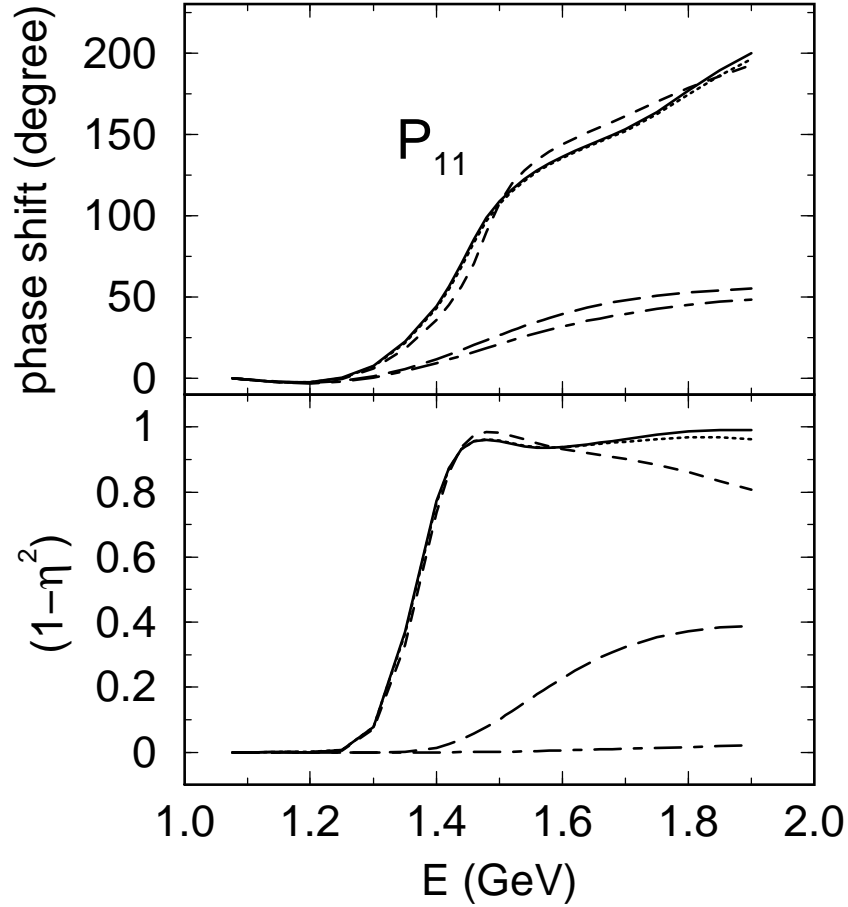,height=12.5cm}
\caption{Phase shift and inelasticity in the partial wave $P_{11}$.
  The curves are calculated using the full model (solid line), the
  channels $\pi N/\sigma N/\pi \Delta$ (dotted line), $\pi N/\pi
  \Delta$ (long dashed line), $\pi N/\sigma N$ (short dashed line),
  and the elastic model (dot-dashed line). The common parameters are 
  the same in all five cases.}
\label{contp11}
\end{center}
\end{figure}

As can be seen in Figs. \ref{phasep11} and \ref{pwa1}, our model
results in a very good description of the $P_{11}$, {\it without}
including a Roper pole diagram.  The rise of the
phase shift and the opening of the inelasticity is generated by the coupling
to the inelastic channels. In Fig. \ref{contp11} we show how the
different reaction channels contribute to the $P_{11}$. The
potential of the elastic model (i.e., where $\pi N$ is the only reaction 
channel) is attractive due to the $\rho$ exchange, and leads to a rising phase
shift without generating a resonant behavior. Including the $\pi
\Delta$ channel hardly improves the situation for the phase shift but
leads to some inelasticity, which starts at about 1.4 GeV. As
soon as we couple to the $\sigma N$ channel, a resonant
shape of the phase shift is generated. The inelasticity opens at 1.3
GeV and reproduces the rapid rise of the experimental data.  Since the
reaction channels $\rho N$ and $\eta N$ scarcely contribute to the
$P_{11}$, decoupling the $\pi \Delta$ channel from the full model
leaves us basically with a $\pi N / \sigma N$ model, which does not
differ much from the full result. Only at higher energies does the $\pi
\Delta $ channel contribute to the inelasticity.

As we have not included a Roper pole diagram into our model, we 
cannot determine any Breit-Wigner parameters from the parameters in
our model.  H\"ohler and Schulte \cite{Hoehler92}, however, were able to
determine resonance parameters from several partial wave solutions by
calculating the speed, which is defined by
\begin{equation}\label{speed}
Sp^{IJLS}(E)=\left|\frac{d\tau^{IJLS}}{dE}\right|,
\end{equation}
and gives some information about the time delay in the reaction
\cite{Goldberger64,Bohm93}.  A resonance causes a large time delay and
will, therefore, form a peak in a diagram in which the speed is plotted
against the energy $E$ (the so-called speed plot). The height and width
of this peak can be related to the mass, width and residue of the
resonance \cite{Hoehler92}.

\begin{figure}[h]
\begin{center}
\epsfig{file=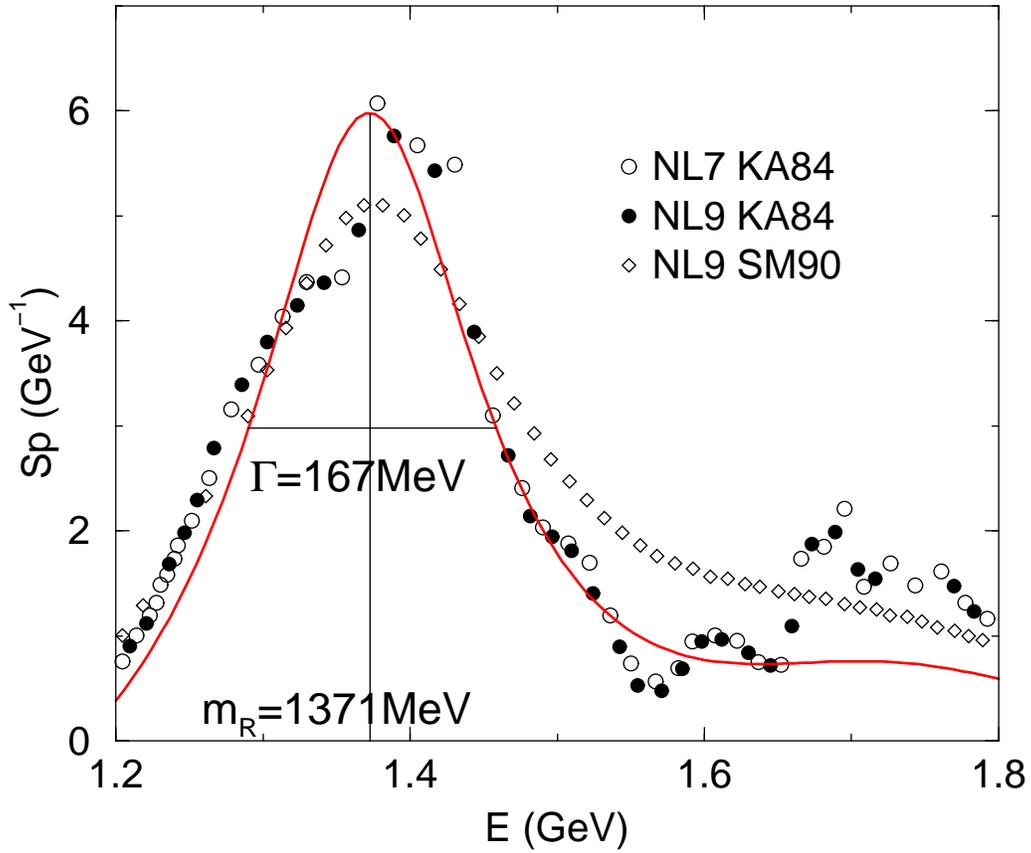,height=11.5cm}
\caption{speed plot in the partial wave $P_{11}$.
The symbols are showing speed plots from Ref. \protect \cite{Hoehler92}
(open circles) and Ref. \protect \cite{Hoehler93} (full circles (KA84 \protect
\cite{KA84,Koch85})
and diamonds (SM90 \protect \cite{SM90})).}
\label{speed_p11}
\end{center}
\end{figure}

The speed plot calculated with our model is displayed in Fig.
\ref{speed_p11}. It agrees very well with the speed plot from the
partial wave solutions KA84\cite{KA84,Koch85} and SM90 \cite{SM90}.
From the height and width we determine the following resonance
parameters (see also Table \ref{tabp11} h):
\begin{eqnarray}
m_R&=&1371 \mbox{ MeV}, \\
\Gamma&=&167 \mbox{ MeV}, \\
r&=&41 \mbox{ MeV}.
\end{eqnarray}
The phase of the residue is lost in taking the absolute value in Eq.
(\ref{speed}) and cannot be determined without making further
assumptions.  In Table \ref{tabp11} our result (i) is compared to the
parameters from the speed plot analyses of H\"ohler and Schulte (f --
h).  The agreement in mass is very good. Besides the width and residue
of the VPI speed plot analysis f), our values agree with the other
speed plot analyses.  The agreement with the pole position of the two
resent VPI solutions \cite{SP98,SM95} is also very good.

\begin{figure}[h]
\begin{center}
\epsfig{file=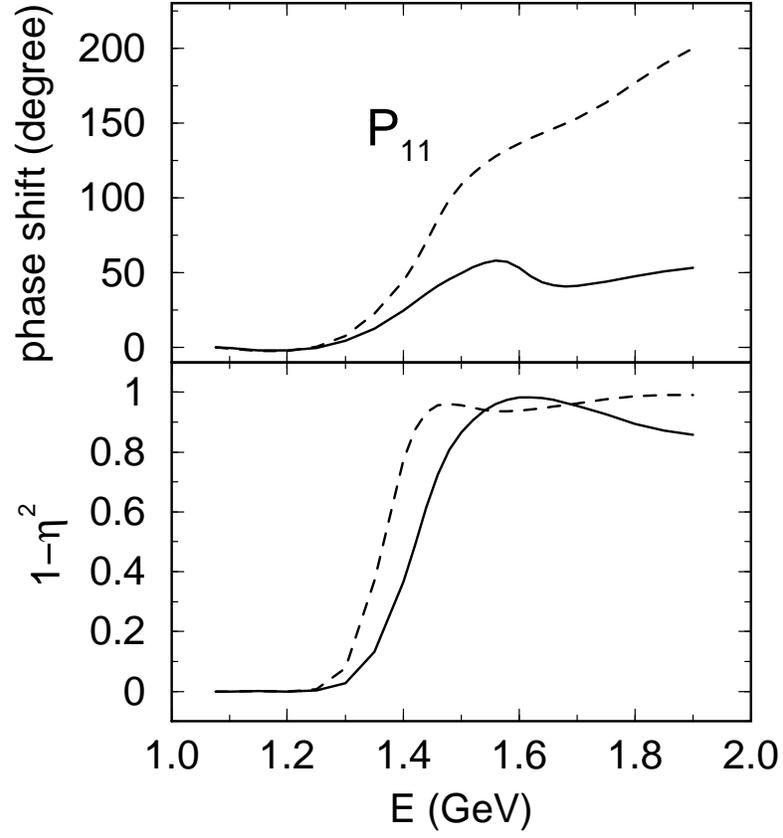,height=11.5cm}
\caption{The partial wave $P_{11}$ calculated with (dashed line) and without
(solid line)
$\pi$ exchange in the $\pi N \to \sigma N$ transition potential, using the same
parameters.}
\label{p11nopit}
\end{center}
\end{figure}

By switching off several contributions in the potential, we have found
the $\pi$ exchange in the transition $\pi N \to \sigma N$ (Fig.
\ref{ropgraf} j) to be very important for the energy dependence of
the $P_{11}$ phase shift. This is demonstrated in Fig. \ref{p11nopit},
where we show the model without $\pi$ exchange in comparison to the
full solution. This contribution is responsible for a large amount of
attraction, especially at higher energies. In contrast, the inelasticity
stays large at higher energies even without $\pi$ exchange, but reaches its
maximum at 1.6 GeV
(the maximum of the full model is located at 1.45 GeV). In an
earlier version of this model \cite{Schuetz98} this contribution was
missing. The attraction that is needed for a good description of the
$P_{11}$ was generated by a strong coupling to the $\sigma N$
channel via the nucleon exchange and a stronger coupling to the $\pi
\Delta$ reaction channel.  However the energy dependence of the $\pi
\Delta$ channel leads to a maximum in the $P_{11}$ phase shift near
1.6 GeV and the phase shift decreases again at higher energies.
Therefore the model \cite{Schuetz98} was restricted to energies below 1.6 GeV.

\begin{figure}[h]
\begin{center}
\epsfig{file=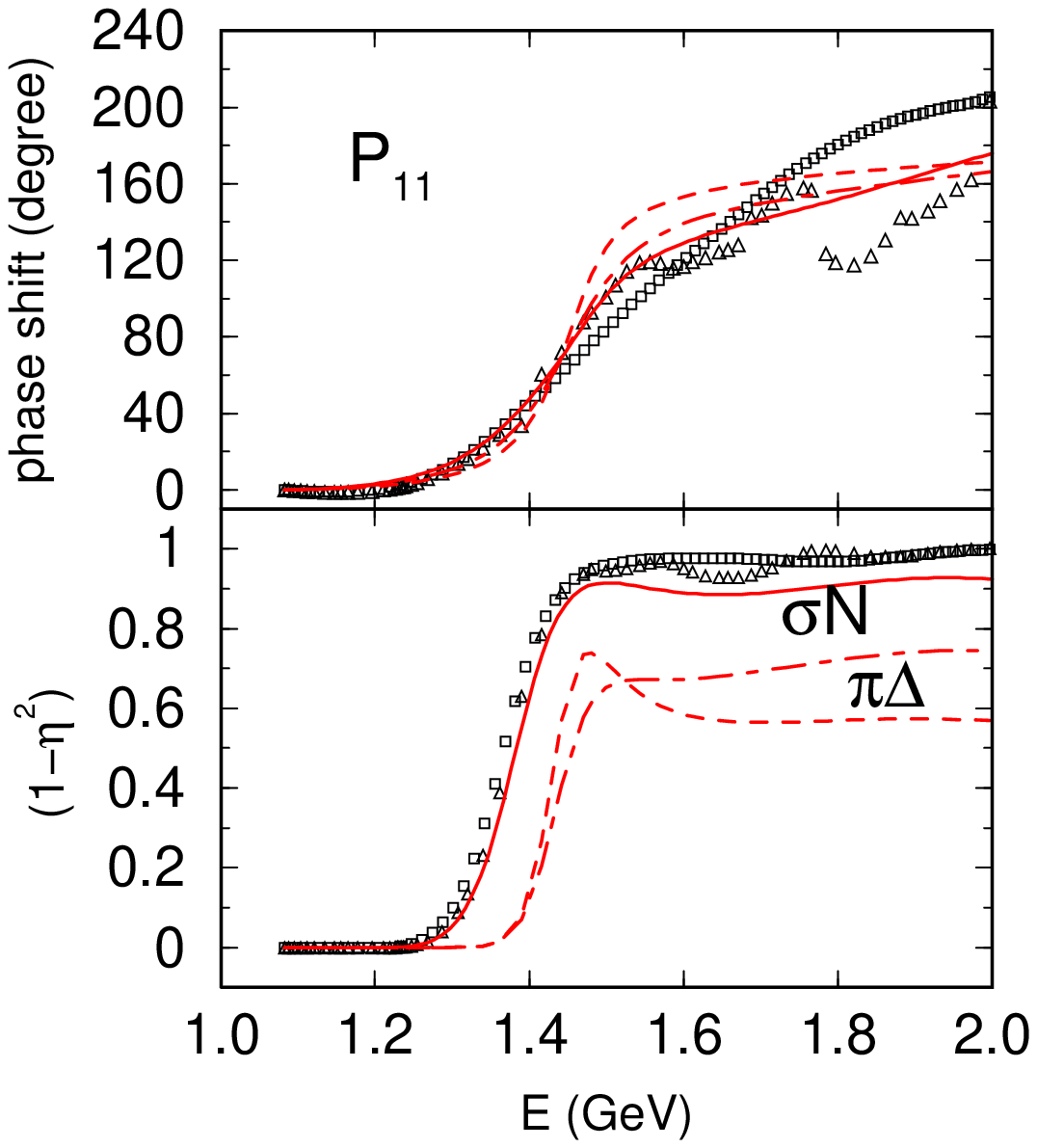,height=11.5cm}
\caption{Results of the simplified model. The solid line was calculated using
  parameter set I of Table \ref{sepparams}, the dashed and dot-dashed curves are
  obtained using sets II and III, respectively. For the solid line only $\pi N$ and
  $\sigma N$ are coupled, whereas for the dashed and dot-dashed lines
  the only channels are $\pi N$ and $\pi \Delta$.}\label{figsep}
\end{center}
\end{figure}

So far we have demonstrated that our model generates a dynamical pole
in the $P_{11}$, which is associated with the Roper resonance. The
phase shift and inelasticity can be described as well as in other
models that include a bare resonance explicitly, and the resonance
parameters from a speed plot analysis are in good agreement with the
speed plot analyses of other partial wave solutions.  We also found,
that the $\sigma N$ and the $\pi \Delta$ channel are important in the
$P_{11}$.  In order to investigate the role of these channels in more
detail, we construct a simplified model that contains the basic
features of the full model used so far. We restrict the simplified
version to the reaction channels $\pi N,\sigma N$ and $\pi\Delta$. A
major simplification is achieved by replacing the microscopic
potential $V_{\mu\nu}(k,k')$ by a separable potential of the
form\footnote{Although the microscopic character of the interaction is
  lost, we can still draw conclusions concerning the role of 
  different reaction channels.}
\begin{equation}
V_{\mu\nu}(k,k')=f_\mu(k)\frac{1}{E-m^0}f_\nu(k'),
\end{equation}
where $m^0$ is a free parameter which (if positive) allows for a pole
in the energy dependence \cite{McLeod85}.  The vertex functions
$f_{\mu}(k)$ are given by
\begin{eqnarray}
f_{N\pi}&=&\sqrt{\frac{3}{8}}\frac{1}{\pi}\frac{f_{N\pi}}{m_{\pi}}k\left(1+
\frac{\omega_{\pi}(k)}{E_N(k)+m_N}\right)N_{\pi N}(k),\\
f_{N\sigma}&=&\frac{g_{N\sigma}}{\sqrt{8}\pi}N_{\sigma N}(k), \\
f_{\Delta\pi}&=&\frac{f_{\Delta\pi}}{m_{\pi}}\frac{k}{\sqrt{6}\pi}\frac{E_{\Delta}(k)\omega_{\pi}(k)}{m_{\Delta}}N_{\pi
\Delta}(k).
\end{eqnarray}
where
$N_{\gamma}(k)=\sqrt{\frac{E_{\gamma}(k)+M_{\gamma}}{E_{\gamma}(k)\omega_{\gamma}(k)}}$.
The coupling constants $f_{N\pi},g_{N\sigma},$ and $f_{\Delta \pi}$ are also
free parameters in the
fit to the $P_{11}$ partial wave amplitude.
All vertex functions are supplemented by a common form factor of the type
(\ref{ffpol})
with a cutoff $\Lambda=2.0$ GeV. The $\pi N$ $T$-matrix can be calculated in
the following way
\cite{Afnan81}:

First we calculate the self-energy
\begin{equation}\label{self}
\Sigma(E)=\sum_{\gamma}\int
q^2dq\frac{|f_{\gamma}(q)|^2}{E-W_{\gamma}-\Sigma_{\gamma}(E_{sub})},
\end{equation}
where the modified propagator (\ref{modprop}) is used for the $\pi \Delta$ and
$\sigma N$ channel.
With this self-energy, the $\pi N$ $T$-matrix can be calculated:
\begin{equation}
T(k',k)=\frac{f_{N\pi}(k')f_{N\pi}(k)}{E-m^0-\Sigma(E)}
\end{equation}
We have fitted the $P_{11}$ phase shift and inelasticity with the three
different sets of parameters shown in Table \ref{sepparams}. Set I only
couples the reaction channels $\pi N$ and $\sigma N$ whereas set II
and III only couple $\pi N$ and $\pi \Delta$. The results for the
different parameter sets are shown in Fig. \ref{figsep}. The $\pi N /
\sigma N$ model describes the $P_{11}$ almost as well as the full
model. In particular, the inelasticity opens at the right energy and the
model results in a continuous rise of the phase shift.  In contrast,
the $\pi N/\pi\Delta$ model (sets II and III) is not able to describe the inelasticity.
The inelastic contributions from the $\pi \Delta$ channel start to
open at higher energies as compared to set I and do not lead to
$(1-\eta^2)\approx 1$. By increasing the coupling to the $\pi
\Delta$ channel (in going from set II to set III) the maximum in the inelasticity can be increased, 
but it still opens at $\approx 1.37$ GeV \footnote{ This
  problem is also present in the separable $\pi N$/$\pi \Delta$ model
  of Blankleider and Walker \cite{Blankleider85}, whereas in the
  separable model of Fuda \cite{Fuda98} the mass of the $\Delta$ is
  adjusted in each partial wave separately in order to describe the
  inelasticies correctly.}. 
So even by increasing the coupling to the $\pi \Delta$ channel, the
onset of inelasticity is not shifted down in energy.
Furthermore the larger coupling (set III)
leads to an overestimation of the phase shift in the energy region of
1.4 -- 1.6 GeV. A good description of the $P_{11}$ partial wave
amplitude with this coupled-channel $\pi N/\pi\Delta$ model is not
possible.

We have also performed a least-squares fit, letting all three coupling
constants and the mass $m^0$ vary freely. The minimizing procedure
always resulted in a negligible coupling to the $\pi \Delta$ channel.
The resulting parameters only differ slightly from the parameter set I
and the curve is almost the same as the solid one in Fig.
\ref{figsep}.

The common feature of the full model discussed at the beginning of
this section and the simplified version introduced here is the use of
the modified propagator (\ref{modprop}) for the $\pi \Delta$ and
$\sigma N$ states, as introduced in Sec. \ref{secmodel}. This allows us
to conclude that a proper treatment of the decay widths of the
intermediate states in the form presented here is very important for
the description of the Roper partial wave.  The self-energy term in the
modified propagator (\ref{modprop}) smears out the threshold of the
$\sigma N$ state over a rather broad energy region. Furthermore it
introduces an additional imaginary part into the amplitude, which
originates from the (energy dependent) decay width of the $\sigma$.
This results in an onset of inelasticity at the correct position. The
strong coupling between the $\pi N$ and the $\sigma N$ channel, as mediated by the 
$t$ channel $\pi$ exchange, generates
large contributions from the rescattering of virtual $\sigma N$ states and
produces the attraction seen in the $P_{11}$.

\section{Summary}
We have presented a coupled-channel model for $\pi N$ scattering in
the energy region from threshold up to 1.9 GeV. The model is based on
an effective Lagrangian and leads to a good description of $\pi N$
partial wave amplitudes.  We have used this model for an investigation of
the Roper resonance. We found, that our full solution of the
relativistic Lippmann-Schwinger equation generates the Roper resonance
dynamically, i.e. without needing a $q^3$ core. We have calculated
resonance parameters by using the speed plot method, and these are consistent
with other analyses.  As source of the dynamical pole we have
identified the $\sigma N$ channel, which we have used together with
the $\pi \Delta$ and $\rho N$ channel as effective description of
$\pi\pi N$ states.  Furthermore we have shown that the most important
features of the our model are the $t$ channel $\pi$ exchange in the $\pi N \to \sigma N$ transition potential and a proper treatment of the decay width of
unstable particles in the quasi-two-body $\pi\pi N$ states.
These results call for a reinvestigation of the Roper resonance in the
quark model, where attention to the role of meson-baryon states, 
or $q^4\bar{q}$ configurations, has to be payed.

\acknowledgments
We are grateful to I.R. Afnan, T. Barnes, and G. H\"ohler for
stimulating discussions.  We would also like to thank J.W.  Durso for
valuable comments while carefully reading the manuscript.  C.H.
acknowledges support from the Alexander-von-Humboldt foundation.  This
work was supported in part by the U.S.  Department of Energy under
Grant No.  DE-FG03-97ER41014.

\appendix

\section{The pseudopotential}\label{ana}
In this appendix we give all expressions for the pseudopotential,
which we use in our coupled-channel model for $\pi N$ scattering.
Let us start with defining some shorthand notation: The on-mass-shell
energies for meson and baryon are
\begin{eqnarray}
\omega_i&=&\sqrt{\vec{p}^2_i+m_i^2}, \nonumber \\
E_i&=&\sqrt{\vec{p}^2_i+m_i^2},
\end{eqnarray}
with the notation as given in Fig. \ref{pingraf}. A common factor
\begin{equation}\label{kappa}
\kappa=\frac{1}{(2\pi)^{3}}\sqrt{\frac{m_1
m_3}{E_1E_3}}\frac{1}{\sqrt{2\omega_2 2\omega_4}}.
\end{equation}
is present in all potentials, which originates from the normalization of fields
and the
relation
\begin{equation}
S_{fi}=\delta_{fi}-2\pi i \delta^4(p_f-p_i) T_{fi}
\end{equation}
between the standard $S$-matrix and the $T$-matrix \cite{Joachain75}.
We use time-ordered perturbation theory (TOPT) in this work
\cite{Schweber62}; therefore all intermediate particles are on their
mass shell (i.e. $p_i^2=m^2_i$ for $i=1..4$). As a consequence the
energy is, in general, not conserved at a vertex, but the total energy
in the reaction, and the three momentum at each vertex, are conserved, as
they must be. In TOPT, a Feynman diagram is represented by two time
orderings (and a possible contact term, which we shall discuss later). The
second time ordering can be constructed out of the first by
replacing the four-momentum $q$ of the intermediate particle with the
momentum $\hat{q}$, which differs only in its 0-th component from $q$:
$\hat{q}^0=-\omega_q$ for meson exchange and $\hat{q}^0=-E_q$ for
baryon exchange. The pseudopotential is then a sum of both time
orders.

The inclusion of the $\Delta$ isobar as an exchanged particle leads to
fundamental difficulties in TOPT. We have therefore chosen the same
pragmatic way of including the $\Delta$ as taken in Refs.
\cite{Schuetz94,Schuetz98}. Since the $\Delta$ exchange contributions
play only a minor role in the investigations of this paper, this
pragmatic approach is justified.

In the following expressions for the pseudopotential, the isospin is
separated. The potentials have to be multiplied by the isospin factors
$IF$, as given in Ref. \cite{Schuetz98}.  Since some contributions -- and the
$\rho N$ channel -- were not included in Ref. \cite{Schuetz98}, we give
the additional relevant isospin factors in Table \ref{tabiso}. The contributions 
can be evaluated in the cm frame by setting $\vec{p}_1=\vec{k}=-\vec{p}_2,
\vec{p}_3=\vec{k}'=-\vec{p}_4$.

The contributions to the pseudopotential
$V^I_{\mu\nu}(\vec{k}',\vec{k},\lambda_1,\lambda_2,\lambda_3,\lambda_4)$
are given by the following expressions:

\subsection{$\npi \rightarrow \npi$}\label{pintopin}
\begin{figure}[h]
\begin{center}
\vspace{5cm}
\epsfig{file=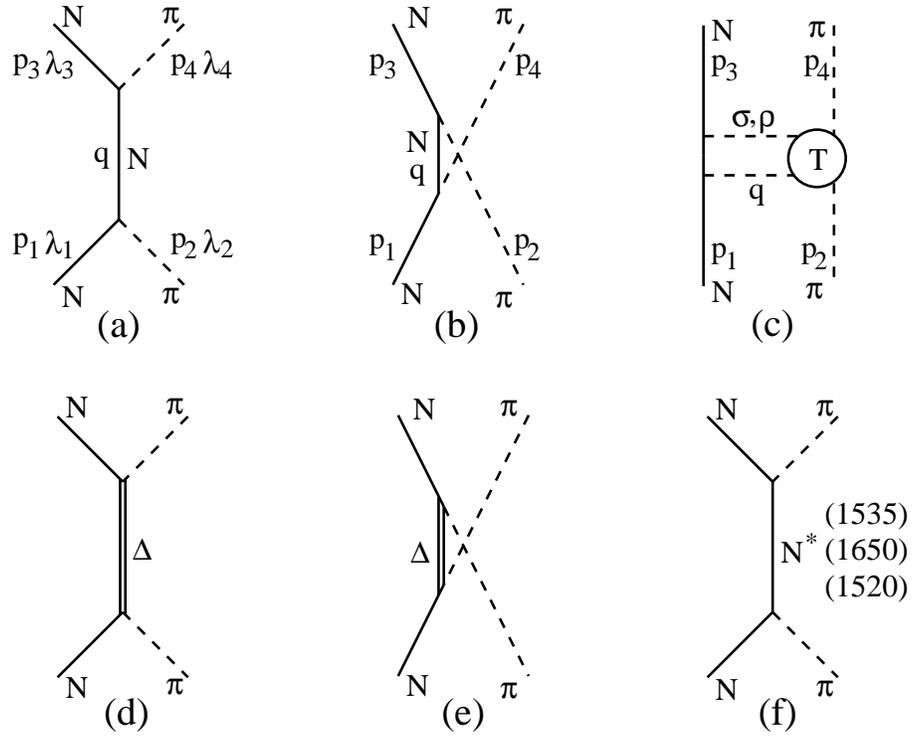,width=12.0cm}
\caption{Contribution to the elastic $\pi N$ interaction.}
\label{pingraf}
\end{center}
\end{figure}

\begin{figure}[h]
\begin{center}
\epsfig{file=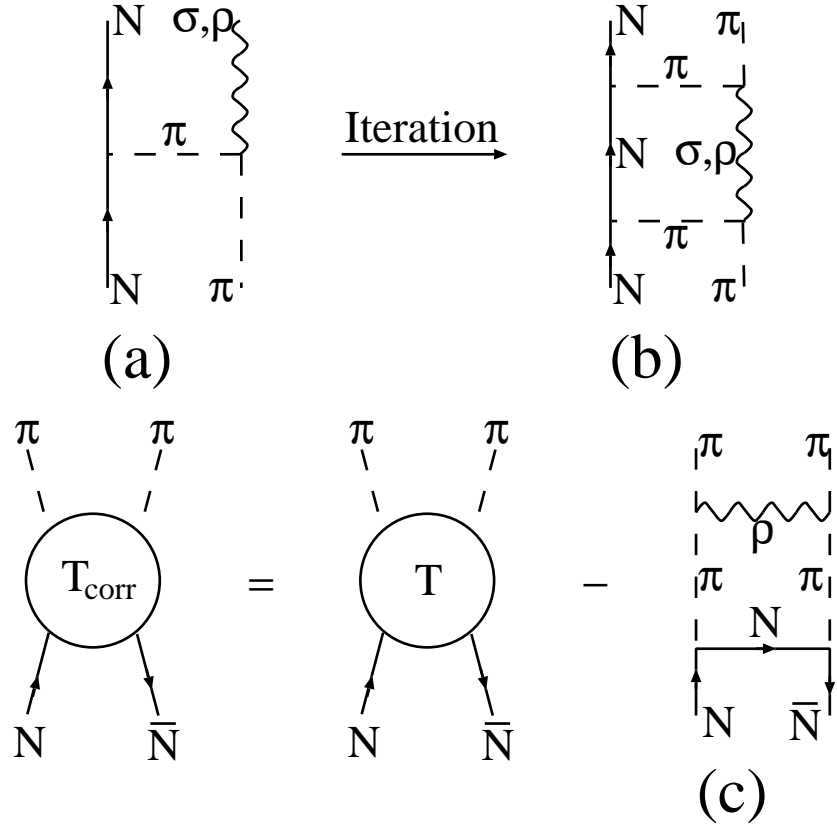,width=11cm}
\caption{Double counting in the correlated $\pi\pi$ exchange arises from iteration of 
the $\pi$ exchange diagram (a), because that generates the box diagram (b), which is
already included
in the correlated $\pi\pi$ exchange (Fig. \ref{pingraf} (c)).
In order to avoid double counting
we remove the diagram (c) from the $N\bar{N}\to\pi\pi$ amplitudes.}
\label{fdouble}
\end{center}
\end{figure}

\begin{itemize}

\item Nucleon pole diagram (Fig. \ref{pingraf}a)
\begin{eqnarray}
\kappa\frac{f^2_{NN\pi}}{m^2_{\pi}} &&
\bar u (\vec{p}_3,\lambda_3)
\gamma_5  p\slas_4 \frac{1}{2m^0_N}
\left(\frac{q\sla+m_N}{E-m^0_N}+
\frac{\hat{q}\sla+m_N}{E-m^0_N-E_1-E_3-\omega_2-\omega_4}\right) \nonumber \\
&&\times\gamma_5  p\slas_2 u(\vec{p}_1,\lambda_1) IF_{Ns}(I).
\end{eqnarray}

\item Nucleon exchange (Fig. \ref{pingraf}b)
\begin{eqnarray}
\kappa\frac{f^2_{NN\pi}}{m^2_{\pi}} &&
\bar u (\vec{p}_3,\lambda_3)
\gamma_5  p\slas_2
\frac{1}{2E_q}\left(\frac{q\sla+m_N}{E-E_q-\omega_2-\omega_4}
+\frac{\hat{q}\sla+m_N}{E-E_q-E_1-E_3}\right)  \nonumber \\
&&\times\gamma_5  p\slas_4 u(\vec{p}_1,\lambda_1) IF_{Nu}(I).
\end{eqnarray}

\item correlated $\pi\pi$ exchange in the {$\sigma$} channel (Fig.
\ref{pingraf}c)\\
\begin{equation}
16\kappa(2p_{2_{\mu}}p_4^{\mu}) \int dt'
\frac{Im(f^0_+(t'))}{(t'-2m_{\pi}^2)(t'-4m_N^2)}
P(t')\bar u (\vec{p}_3,\lambda_3)
u(\vec{p}_1,\lambda_1) IF_{\sigma t}(I),
\end{equation}
where
$P(t')=\frac{1}{2\omega_{t'}}\left(\frac{1}{E-\omega_2-E_3-\omega_{t'}}+\frac{1}{E-\omega_4-E_1-\omega_{t'}}\right)$,
$\omega_{t'}=\sqrt{q^2+t'}$ and $f$ is a Frazer-Fulco amplitude
\cite{Frazer60,Krehl99}.

\item correlated $\pi\pi$ exchange in the {$\rho$} channel (Fig. \ref{pingraf}c)\\
\begin{eqnarray}
-12\kappa&&\left[ \frac{Q^{\mu}(P_1+P_3)_{\mu}}{2m_N}
\int dt' Im(\Gamma_2(t'))P(t') \bar u (\vec{p}_3,\lambda_3)
u(\vec{p}_1,\lambda_1) \right. \nonumber \\
&& \left. -\int dt'Im(\Gamma_2(t')+\Gamma_1(t'))P(t') \bar u
(\vec{p}_3,\lambda_3) Q\hspace{-0.07cm}\sla
\hspace{0.07cm}u(\vec{p}_1,\lambda_1)
\right] IF_{\rho t}(I),
\end{eqnarray}
where
$\Gamma_1(t)=-\frac{m_N}{p_t^2}\left(f^1_+(t)-\frac{t}{4\sqrt{2}m_N}f^1_-(t)\right)$,
$\Gamma_2(t)=\frac{m_N}{p_t^2}\left(f^1_+(t)-\frac{m_N}{\sqrt{2}}f^1_-(t)\right)$,
and $Q=\frac{1}{2}(p_2+p_4)$.

\item {$\Delta$} pole diagram (Fig. \ref{pingraf}d)
\begin{equation}
\kappa\frac{f^2_{N\Delta\pi}}{m^2_{\pi}}
\bar u (\vec{p}_3,\lambda_3) p_{4_\mu}
\frac{P^{\mu\nu}(q)}{(E-m_{\Delta})(E+m_{\Delta})}
p_{2_\nu}
u(\vec{p}_1,\lambda_1) IF_{\Delta s}(I).
\end{equation}

\item {$\Delta$} exchange (Fig. \ref{pingraf}f)
\begin{eqnarray}
\kappa\frac{f^2_{N\Delta\pi}}{m^2_{\pi}}&&
\bar u (\vec{p}_3,\lambda_3) p_{2_\mu} P^{\mu\nu}(q)
\left(\frac{1}{2E_q(E-E_q-\omega_2-\omega_4)}+
\frac{1}{2E_q(E-E_q-E_1-E_3)}\right) \nonumber \\
&&\times p_{4_\nu}
u(\vec{p}_1,\lambda_1) IF_{\Delta u}(I).
\end{eqnarray}

\item $N^*(S_{11})$ pole diagram (Fig. \ref{pingraf}g)
\begin{equation}\label{n*s11}
\kappa g^2_{N^*N\pi}
\bar u (\vec{p}_3,\lambda_3)
\frac{1}{2m^0_{N^*}}
\frac{q\sla+m^0_N}{E-m^0_{N^*}}
u(\vec{p}_1,\lambda_1) IF_{N^*s}(I).
\end{equation}

\item $N^*(D_{13})$ pole diagram (Fig. \ref{pingraf}g)
\begin{equation}\label{n*d13}
\kappa \frac{f^2_{N^*N\pi}}{m_{\pi}^4}
\bar u (\vec{p}_3,\lambda_3)
\gamma^5 p\slas_4 p_{4_\mu}
\frac{1}{2m^0_{N^*}}
\frac{P^{\mu\nu}(q)}{E-m^0_{N^*}}
\gamma^5 p\slas_2 p_{2_\nu}
u(\vec{p}_1,\lambda_1) IF_{N^*s}(I).
\end{equation}
\end{itemize}

The tensor $P^{\mu\nu}$ is given by
\begin{equation}
P^{\mu\nu}(p)=(p\sla +M)
\left[-g^{\mu\nu}+\frac{1}{3}\gamma^{\mu}\gamma^{\nu}+
\frac{2}{3M^2}p^{\mu}p^{\nu}-\frac{1}{3M}(p^{\mu}\gamma^{\nu}-p^{\nu}\gamma^{\mu})\right],
\end{equation}
where $M$ is the mass of the exchanged baryon.

\subsection{$\npi \rightarrow \nrho$}\label{pintorhon}
\begin{figure}[h]
\begin{center}
\vspace{1.5cm}
\epsfig{file=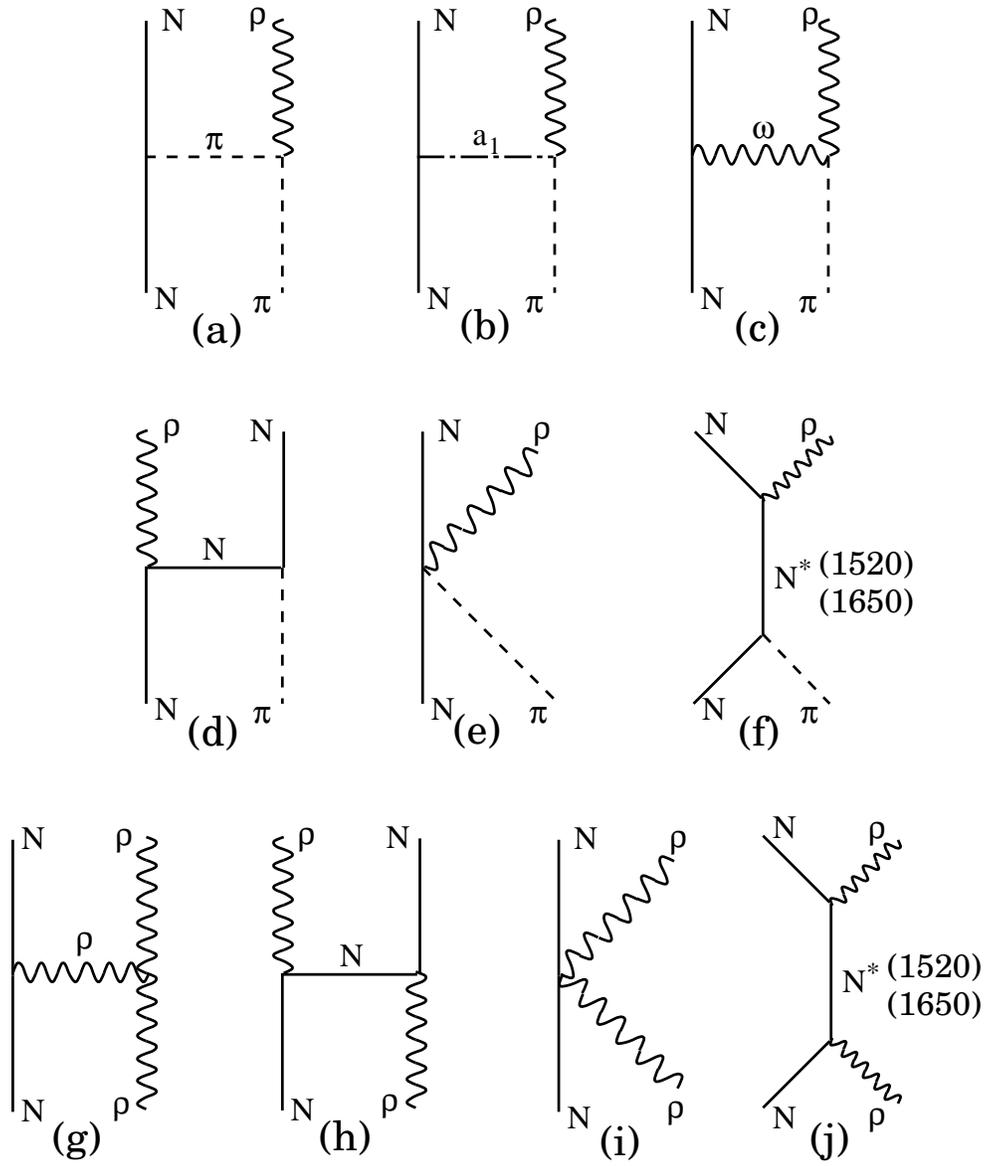,width=13cm}
\caption{The potential for the coupling to the $\rho N$ channel.}
\label{rhongraf}
\end{center}
\end{figure}

\begin{itemize}

\item {$\pi$} exchange (Fig. \ref{rhongraf}a)
\begin{eqnarray}\label{piex}
-\kappa g_{\rho\pi\pi}\frac{f_{NN\pi}}{m_{\pi}}&&
\bar u (\vec{p}_3,\lambda_3) \gamma^5\gamma^{\mu}
u(\vec{p}_1,\lambda_1)\left(\frac{q_{\mu}(p_2-q)_{\nu}}{2\omega_q(E-\omega_q-E_3-\omega_2)}\right.
\nonumber \\
&&\left.
+\frac{\hat{q}_{\mu}(p_2-\hat{q})_{\nu}}{2\omega_q(E-\omega_q-E_1-\omega_4)}\right)
\epsilon^{*,\nu}(\vec{p}_4,\lambda_4) IF_{\pi}(I),
\end{eqnarray}
where $\epsilon^{\nu}(\vec{p}_4,\lambda_4)$ is the polarisation vector
of a massive spin 1 particle with momentum $p_4$ and helicity
$\lambda_4$ \cite{GrossBook}.

\item {$a_1$} exchange (Fig. \ref{rhongraf}b)
\begin{eqnarray}\label{a1ex}
2\kappa g_{\rho}\frac{f_{NN\pi}}{m_{\pi}}&&
\bar u (\vec{p}_3,\lambda_3) \gamma^5 \gamma_{\mu} u(\vec{p}_1,\lambda_1)
\left(\frac{-g^{\mu\nu}+\frac{q^{\mu}q^{\nu}}{m_{a_1}^2}}{2\omega_q(E-\omega_q-E_3-\omega_2)}\right.
 \nonumber \\
&&\times \left.[(p_2+\frac{q}{2})_{\tau}
p^{\tau}_4\epsilon^*_{\nu}(\vec{p}_4,\lambda_4)-(p_2+\frac{q}{2})^{\tau}\epsilon_{\tau}^*(\vec{p}_4,\lambda_4)
p_{4_\nu}]\right. \nonumber \\
&&+[(p_2+\hat{q}/2)_{\tau} p^{\tau}_4\epsilon^*_{\nu}(\vec{p}_4,\lambda_4)-
(p_2+\hat{q}/2)^{\tau}\epsilon_{\tau}^*(\vec{p}_4,\lambda_4) p_{4_\nu}]
 \nonumber \\
&&\times
\left.\frac{-g^{\mu\nu}+\frac{\hat{q}^{\mu}\hat{q}^{\nu}}{m_{a_1}^2}}{2\omega_q(E-\omega_q-E_1-\omega_4)}\right)
IF_{a_1}(I).
\end{eqnarray}

\item {$\omega$} exchange (Fig. \ref{rhongraf}c)
\begin{eqnarray}
\kappa g_{NN\omega}\frac{g_{\omega\pi\rho}}{m_{\omega}}&&
\bar u (\vec{p}_3,\lambda_3)\left(
[\gamma^{\tau}-i\frac{\kappa_{\omega}}{2m_N} \sigma^{\tau\nu} q_{\nu}]
\frac{1}{2\omega_q(E-\omega_q-E_3-\omega_2)}\right. \nonumber \\
&&\left.+[\gamma^{\tau}-i\frac{\kappa_{\omega}}{2m_N} \sigma^{\tau\nu}
\hat{q}_{\nu}]
\frac{1}{2\omega_q(E-\omega_q-E_1-\omega_4)}\right) \nonumber \\
&&\times u(\vec{p}_1,\lambda_1)
\epsilon_{\mu\alpha\lambda\tau}p^{\alpha}_4\epsilon^{*,\mu}(\vec{p}_4,\lambda_4)p^{\lambda}_2
IF_{\omega}(I),
\end{eqnarray}
with $\epsilon_{0123}=-1$.
\item Nucleon exchange (Fig. \ref{rhongraf}d)
\begin{eqnarray}
-i \kappa g_{NN\rho}\frac{f_{NN\pi}}{m_{\pi}} &&
\bar u (\vec{p}_3,\lambda_3)
\gamma_5  p\slas_2
\left(\frac{q\sla+m_N}{E-E_q-\omega_2-\omega_4}+
\frac{\hat{q}\sla+m_N}{E-E_q-E_1-E_3}\right) \nonumber \\
&&\times \frac{1}{2E_q} [\epsilon\sla^*(\vec{p}_4,\lambda_4)
-i\frac{\kappa_{\rho}}{2m_N}\sigma^{\mu\nu}p_{4_\nu}\epsilon^*_{\mu}(\vec{p}_4,\lambda_4)]
u(\vec{p}_1,\lambda_1) IF_{Nu}(I).
\end{eqnarray}

\item {$NN\pi\rho$} contact graph (Fig. \ref{rhongraf}e)
\begin{equation}
-\kappa g_{\rho}\frac{f_{NN\pi}}{m_{\pi}}
\bar u (\vec{p}_3,\lambda_3)
\gamma^5 \epsilon\sla^*(\vec{p}_4,\lambda_4)
u(\vec{p}_1,\lambda_1) IF_{ct}(I).
\end{equation}

\item {$N^*(S_{11})$} pole diagram (Fig. \ref{rhongraf}f)
\begin{eqnarray}
\kappa g_{N^*N\rho} g_{N^*N\pi} &&
\bar u (\vec{p}_3,\lambda_3)
\gamma^5[\gamma^{\mu}-i\frac{\kappa_{N^*N\rho}}{2m_{N^*}}\sigma^{\mu\nu}p_{4_\nu}]\epsilon^*_{\mu}(\vec{p}_4,\lambda_4)
\nonumber \\
&&\times \frac{1}{2m^0_{N^*}}\frac{q\sla+m^0_{N^*}}{E-m^0_{N^*}}
u(\vec{p}_1,\lambda_1) IF_{N^*s}(I).
\end{eqnarray}

\item {$N^*(D_{13})$} pole diagram (Fig. \ref{rhongraf}f)
\begin{eqnarray}
i\kappa\frac{f_{N^*N\pi}f_{N^*N\rho}}{m^2_{\pi}m_{\rho}} &&
\bar u (\vec{p}_3,\lambda_3)
(p\slas_{4}\epsilon^*(\vec{p}_4,\lambda_4)-p_{4_\mu}\epsilon\sla^*(\vec{p}_4,\lambda_4))
\nonumber \\
&&\times \frac{P^{\mu\nu}(q)}{2m^0_{N^*}(E-m^0_{N^*})}
p_{2_\nu} \gamma^5 p\slas_2
u(\vec{p}_1,\lambda_1) IF_{N^*s}(I).
\end{eqnarray}
\end{itemize}

Since we are using time ordered perturbation theory \cite{Schweber62}, which
is a formalism based on the Hamiltonian instead of the Lagrangian, we
must transform the Lagrangian to the Hamiltonian via the Legendre
transformation
\begin{equation}\label{legtrafo}
{\cal H}=\sum_{j}\frac{\delta{\cal L}}{\delta\dot{\Phi}_j}\dot{\Phi}_j -{\cal
L},
\end{equation}
where $\Phi_j$ are the fields in ${\cal L}$.
This transformation introduces additional terms into the interaction,
which, in our case, are of the form of contact interactions \cite{Bockmann99}.
In TOPT all particles are on the mass shell, so that the 0-th
component of the exchanged particle, ($q^0=\sqrt{\vec{q}^2+m_{X}^2}$),
is quite different from the one in covariant perturbation theory (e.g.,
$q^0=p_1^0-p_3^0$ for a $t$-channel exchange). Therefore the potential is
different in the two approaches as soon as a time derivative acts on the
filed of the exchanged particle. Since both approaches ultimately must lead to the
same on-shell potential, the role of the additional interactions is to restore
the equivalence between TOPT and covariant perturbation theory \cite{Bockmann99}.

Since both the $\pi NN$ and the $\pi\rho a_1$ Lagrangians contain a
time derivative on the $\pi$ and the $a_1$, there are
additional terms for the $\pi $ and the $a_1$ exchange
contributions, which have to be added to Eqs. (\ref{piex}) and (\ref{a1ex}),
respectively .
For $\pi$ exchange this term is
\begin{equation}
\kappa g_{\rho}\frac{f_{\pi NN}}{m_{\pi}}
\bar u (\vec{p}_3,\lambda_3) \gamma^5\gamma^{0}
u(\vec{p}_1,\lambda_1)\epsilon_0^{*}(\vec{p}_4,\lambda_4) IF_{\pi}(I),
\end{equation}
and for $a_1$ exchange it is
\begin{equation}
2\kappa g_{\rho}\frac{f_{\pi NN}}{m_{\pi}}
\frac{1}{m_{a_1}^2}\bar u (\vec{p}_3,\lambda_3) \gamma^5 \gamma_{0}
u(\vec{p}_1,\lambda_1)
[{p_2}_{\mu}
p^{\mu}_4\epsilon^*_{0}(\vec{p}_4,\lambda_4)-p^{\mu}_2\epsilon_{\mu}^*(\vec{p}_4,\lambda_4)
p_{4_0}]
IF_{a_1}(I).
\end{equation}

\subsection{$\nrho \rightarrow \nrho$}\label{rhontorhon}

\begin{itemize}
\item {$\rho$} exchange (Fig. \ref{rhongraf} g)
\begin{eqnarray}\label{rhoex}
-i\kappa \frac{g^2_{\rho}}{2} &&
\left(\bar{u} (\vec{p}_3,\lambda_3)
[\gamma^{\mu}-i\frac{\kappa_{\rho}}{2m_N} \sigma^{\mu\nu} q_{\nu}]
u (\vec{p}_1,\lambda_1)
\frac{1}{2\omega_q(E-\omega_q-E_3-\omega_2)}\right. \nonumber \\
&&\times [\epsilon^{\tau}(\vec{p}_2,\lambda_2)
\epsilon_{\tau}^*(\vec{p}_4,\lambda_4)(-p_4-p_2)_{\mu}+
(q+p_4)^{\tau} \epsilon_{\tau}(\vec{p}_2,\lambda_2)
\epsilon^*_{\mu}(\vec{p}_4,\lambda_4)+ \nonumber \\
&&(p_2-q)^{\tau} \epsilon_{\tau}^*(\vec{p}_4,\lambda_4)
\epsilon_{\mu}(\vec{p}_2,\lambda_2)] \nonumber \\
&&+[\epsilon^{\tau}(\vec{p}_2,\lambda_2)
\epsilon_{\tau}^*(\vec{p}_4,\lambda_4)(-p_4-p_2)_{\mu}+
(\hat{q}+p_4)^{\tau} \epsilon_{\tau}(\vec{p}_2,\lambda_2)
\epsilon^*_{\mu}(\vec{p}_4,\lambda_4)+ \nonumber \\
&&(p_2-\hat{q})^{\tau} \epsilon_{\tau}^*(\vec{p}_4,\lambda_4)
\epsilon_{\mu}(\vec{p}_2,\lambda_2)] \nonumber \\
&&\times \left.\bar{u} (\vec{p}_3,\lambda_3)
[\gamma^{\mu}-i\frac{\kappa_{\rho}}{2m_N} \sigma^{\mu\nu} \hat{q}_{\nu}]
u (\vec{p}_1,\lambda_1)
\frac{1}{2\omega_q(E-\omega_q-E_1-\omega_4)} \right) IF_{\rho}(I).
\end{eqnarray}

\item Nucleon exchange (Fig. \ref{rhongraf} h)
\begin{eqnarray}
\kappa g^2_{NN\rho} &&
\bar u (\vec{p}_3,\lambda_3)
[\gamma^{\mu}+i\frac{\kappa_{\rho}}{2m_N} \sigma^{\mu\nu} p_{2_\nu}]
\epsilon_{\mu}(\vec{p}_2,\lambda_2) \nonumber \\
&&\times \frac{1}{2E_q}
\left(\frac{q\sla+m_N}{E-E_q-\omega_2-\omega_4}+
\frac{\hat{q}\sla+m_N}{E-E_q-E_1-E_3}\right) \nonumber \\
&&\times [\gamma^{\tau}-i\frac{\kappa_{\rho}}{2m_N} \sigma^{\tau\nu} p_{4_\nu}]
\epsilon^*_{\tau}(\vec{p}_4,\lambda_4)
u(\vec{p}_1,\lambda_1) IF_{Nu}(I).
\end{eqnarray}

\item {$NN\rho\rho$} contact graph (Fig. \ref{rhongraf} i)
\begin{equation}
\kappa g_{NN\rho}\frac{f_{NN\rho}}{2m_N}
\bar u (\vec{p}_3,\lambda_3)
\sigma^{\mu\nu}
\epsilon_{\mu}(\vec{p}_2,\lambda_2)\epsilon^*_{\nu}(\vec{p}_4,\lambda_4)
u(\vec{p}_1,\lambda_1) IF_{ct}(I).
\end{equation}

\item {$N^*(S_{11})$} pole diagram (Fig. \ref{rhongraf} j)
\begin{eqnarray}
\kappa g^2_{N^*N\rho} &&
\bar u (\vec{p}_3,\lambda_3)
\gamma^5[\gamma^{\mu}-i\frac{\kappa_{N^*N\rho}}{2m_{N^*}}\sigma^{\mu\nu}p_{4_\nu}]\epsilon^*_{\mu}(\vec{p}_4,\lambda_4)
\frac{1}{2m^0_{N^*}}\frac{q\sla+m^0_{N^*}}{E-m^0_{N^*}} \nonumber \\
&&\times
\gamma^5[\gamma^{\mu}+i\frac{\kappa_{N^*N\rho}}{2m_{N^*}}\sigma^{\mu\nu}p_{2_\nu}]\epsilon_{\mu}(\vec{p}_2,\lambda_2)
u(\vec{p}_1,\lambda_1) IF_{N^*s}(I).
\end{eqnarray}

\item {$N^*(D_{13})$} pole diagram
\begin{eqnarray}
\kappa \frac{f^2_{N^*N\rho}}{m^2_{\rho}} &&
\bar{u} (\vec{p}_3,\lambda_3)
(p\slas_4\epsilon^*_{\mu}(\vec{p}_4,\lambda_4)-p_{4_\mu}\epsilon\sla^*(\vec{p}_4,\lambda_4))
\frac{P^{\mu\nu}(q)}{2m^0_{N^*}(E-m^0_{N^*})} \nonumber \\
&&\times
(p\slas_2\epsilon_{\nu}(\vec{p}_2,\lambda_2)-p_{2_\nu}\epsilon\sla(\vec{p}_2,\lambda_2))
u (\vec{p}_1,\lambda_1)
IF_{N^*s}(I).
\end{eqnarray}
\end{itemize}

The $\rho NN$ coupling from Table \ref{tablag} contains a time derivative
of the $\rho$ field, which causes an additional term in the Hamiltonian.
On-shell, this term cancels the $q^{\mu}q^{\nu}$ term of the spin-1 propagator,
which is
also approximately true off-shell. Therefore we can mimic the additional
contact term in TOPT
by using the reduced spin-1 propagator,
\begin{equation}\label{effprop}
\frac{-g^{\mu\nu}}{E-\omega_q-E_3-\omega_2}
+\frac{-g^{\mu\nu}}{E-\omega_q-E_1-\omega_4}.
\end{equation}
We have checked numerically that the exact procedure leads only
to tiny differences in the off-shell potential. We have applied this reduced
spin-1 propagator
to the $\rho$ exchange contribution (\ref{rhoex}) above.

\subsection{$\npi \rightarrow \dpi$}\label{pintopidelta}
Due to relative signs in our Lagrangian (Table \ref{tablag}), the nucleon,
$\Delta$ and
$\rho$ exchange contributions from Ref. \cite{Schuetz98} must be multiplied
by a minus sign.
In addition, we have included the {$N^*(D_{13})$} pole diagram (Fig.
\ref{ropgraf}d):
\begin{equation}
-\kappa \frac{f_{N^*N\pi}f_{N^*\Delta\pi}}{m^3_{\pi}}
\bar{u}_{\mu} (\vec{p}_3,\lambda_3) p\slas_4
\frac{P^{\mu\nu}(q)}{2m^0_{N^*}}
\frac{1}{E-m^0_{N^*}}
p_2^{\nu} \gamma^5 p\slas_2
u(\vec{p}_1,\lambda_1) IF_{N^*s}(I).
\end{equation}

\subsection{$\dpi \rightarrow \dpi$}\label{pideltatopidelta}
\begin{figure}[h]
\begin{center}
\vspace{1.5cm}
\epsfig{file=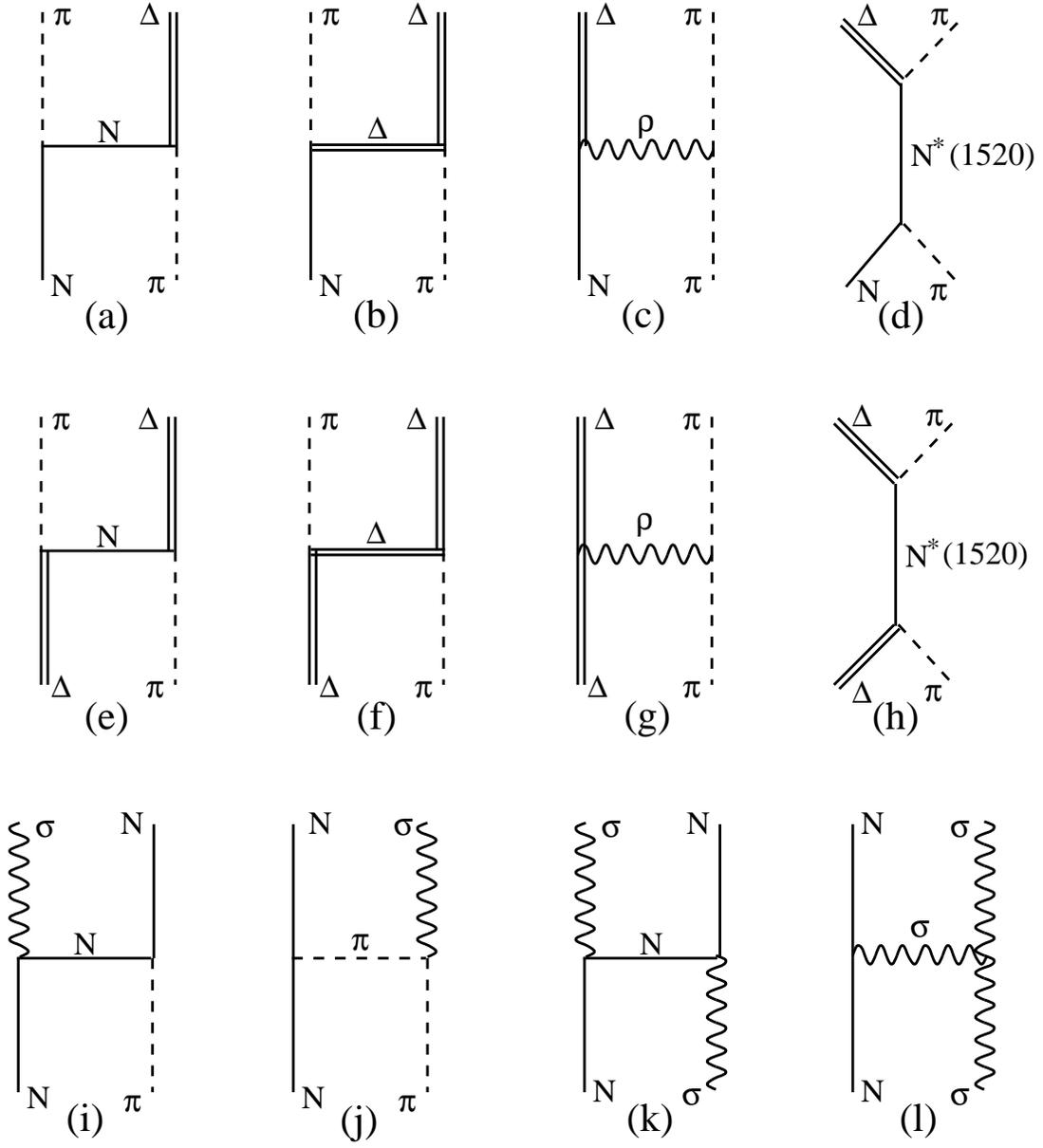,width=14.5cm}
\caption{Additional diagrams for coupling to the  $\pi \Delta$ and $\sigma N$
channels.}
\label{ropgraf}
\end{center}
\end{figure}

The nucleon and $\Delta$ exchange can be taken from Ref. \cite{Schuetz98}. Here 
we do not
use a Gordon decomposition for the $\rho$ exchange (Fig. \ref{ropgraf}g),
which therefore has the form
\begin{eqnarray}
i\kappa g_{\Delta\Delta\rho}g_{\rho\pi\pi} &&
\bar{u}^{\tau} (\vec{p}_3,\lambda_3)
\left[\gamma_{\mu}-i\frac{\kappa_{\Delta\Delta\rho}}{2m_{\Delta}}\sigma_{\mu\nu}q^{\nu}\right]
u_{\tau}(\vec{p}_1,\lambda_1)
\frac{1}{2\omega_q} \nonumber \\
&&\times
\left(\frac{1}{E-\omega_q-\omega_2-E_3}+\frac{1}{E-\omega_q-\omega_4-E_1}\right)
(p_2+p_4)^{\mu} IF_{\rho}(I),
\end{eqnarray}
and we have used the reduced spin-1 propagator from Eq. (\ref{effprop}).
We have also included the {$N^*(D_{13})$} pole diagram (Fig.
\ref{ropgraf}h)
\begin{equation}
\kappa \frac{f^2_{N^*\Delta\pi}}{m_{\pi}^2}
\bar{u}_{\mu} (\vec{p}_3,\lambda_3) p\slas_4
\frac{P^{\mu\nu}(q)}{2m^0_{N^*}} \frac{1}{E-m^0_{N^*}}
p\slas_2
u_{\nu}(\vec{p}_1,\lambda_1) IF_{N^*s}(I).
\end{equation}

\subsection{$\npi \rightarrow \nsig$ and $\nsig \rightarrow
\nsig$}\label{pintosigman}
We take over the contributions from Ref. \cite{Schuetz98}, but additionally
use a $\pi$ exchange contribution for the $\pi N \to \sigma N$ transition (Fig.
\ref{ropgraf}j)
\begin{eqnarray}\label{piaus}
i \kappa \frac{f_{NN\pi}}{m_{\pi}}\frac{g_{\sigma \pi \pi}}{m_{\pi}}&&
\bar{u} (\vec{p}_3,\lambda_3) \frac{1}{2\omega_q}
\left( \frac{\gamma^5 q\sla q_{\mu}}{E-E_q-\omega_2-E_3}\right. \nonumber \\
&&\left.+
\frac{\gamma^5 \hat{q}\sla \hat{q}_{\mu}}{E-E_q-\omega_4-E_1}\right)
p_2^{\mu}
u(\vec{p}_1,\lambda_1) IF_{\pi}(I),
\end{eqnarray}
which again must be supplemented by the additional term
\begin{equation}\label{pict}
i \kappa \frac{f_{\pi NN}}{m_{\pi}}\frac{g_{\sigma \pi \pi}}{m_{\pi}}
\bar{u}(\vec{p}_3,\lambda_3) \gamma^5 \gamma^0 p_2^0 u(\vec{p}_1,\lambda_1)
IF_{\pi}(I)
\end{equation}
resulting from the
Legendre transformation (\ref{legtrafo}).

\subsection{The $\neta$ reaction channel}
\begin{figure}[h]
\begin{center}
\vspace{5cm}
\epsfig{file=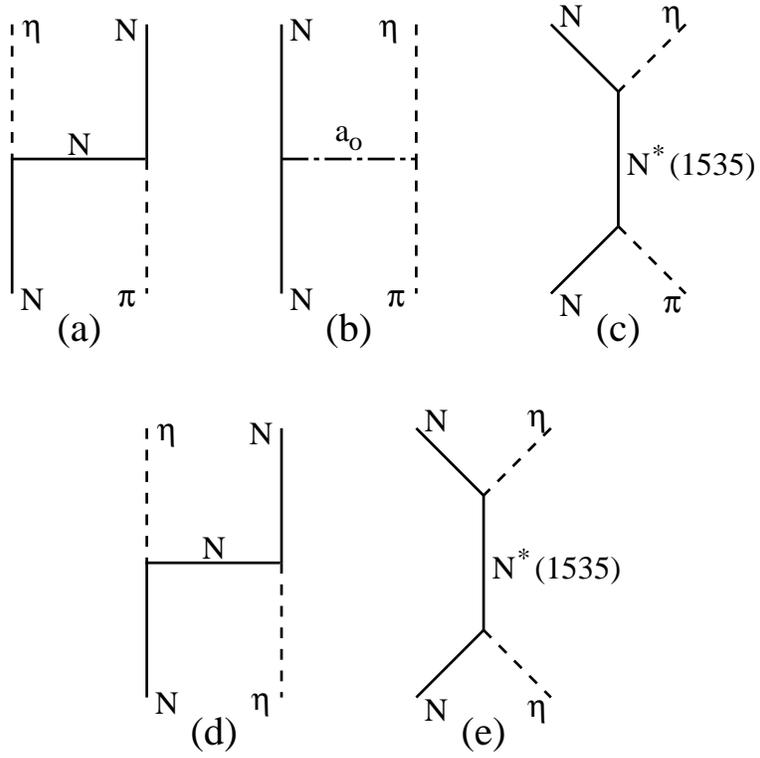,width=10.0cm}
\caption{Additional contribution in coupling to the $\eta N$ channel.}
\label{etangraf}
\end{center}
\end{figure}

The coupling to the $\eta N$ channel (Fig. \ref{etangraf}) can be taken from
Ref.
\cite{Schuetz98}. The additional coupling of the $N_{D_{13}}^*(1520)$
can be constructed from the $D_{13}$ pole diagram of the direct $\pi N$
interaction by replacing one ($\pi N \to \eta N$) or two (direct $\eta
N$) $N^*N\pi$ coupling constants by the $N^*N\eta$ coupling, respectively.

\newpage


\begin{table}
\begin{center}
\begin{tabular}{|cllccl|}
&$m_R$ & $\Gamma$& pole  & residue ($r$,$\phi$)&  Ref.\\
&(MeV) & (MeV)   & (MeV)& $r$ in MeV,$\phi$ in $^o$ &\\
\hline
a)&1467 & 440 & $1346-i88$ & (42,-101) &\cite{SM95}\\
b)&1456 & 428 & $1361-i86$ & (36,-78) &\cite{SP98}\\
c)&1462(10) & 391(34) & -- & -- & \cite{Manley92}\\
d)&1471 & 545 & $1370-i114$ & (74,-84)& \cite{Cutkosky90} \\
e)&1479 & -- & $1383-i158$ & -- & \cite{Vrana99} \\
\hline
f)&1375(30) & 180(40) & -- & (52(5),-100(35)) &\cite{Hoehler92} CMB\\
g)&1360 & 252 & -- & (109,-93) &\cite{Hoehler92} VPI\\
h)&1385(9) & 164(35) & -- & (40,--) &\cite{Hoehler92} KA\\
i)&1371 & 167 & -- & (41,--) & this work \\
\end{tabular}
\end{center}
\caption{Some analyses of the $\pi N$ partial wave $P_{11}$ as listed in the
  Review of Particle Physics \protect \cite{PDG98}.  The resonance
  parameters are denoted by $m_R$ for the mass and $\Gamma$ for the
  width of the resonance. The residue is parameterized by $re^{i\phi}$.
  The numbers in brackets give the error in the last digit. For
  analyses f),g), and h) the abbreviations CMB \protect \cite{Cutkosky79},
  VPI \protect \cite{SM90} and KA\protect \cite{Koch85} indicate for
  which partial wave solution the speed plot is calculated.}
\label{tabp11}
\end{table}

\begin{table}
\begin{center}
\begin{tabular}{|l|l|}
Vertex & ${\cal L}_{int}$ \\
\hline
$NN\pi$
&$-\frac{f_{NN\pi}}{m_{\pi}} \bar{\Psi}\gamma^5\gamma^{\mu}
\vec{\tau}\partial_{\mu}\vec{\pi} \Psi $ \\
$N\Delta\pi$
&$\frac{f_{N \Delta\pi}}{m_{\pi}} \bar{\Delta}^{\mu} \vec{T}^{\dagger}
\partial_{\mu} \vec{\pi} \Psi + \text{h.c.}$ \\
$\rho\pi\pi$
&$-g_{\rho \pi \pi} (\vec{\pi} \times \partial_{\mu} \vec{\pi})
\vec{\rho}^{\mu} $\\
$NN\rho$
&$-g_{NN\rho}
\bar{\Psi}[\gamma^{\mu}-\frac{\kappa_{\rho}}{2m_N}\sigma^{\mu\nu}\partial_{\nu}]\vec{\tau}\vec{\rho}_{\mu}
\Psi $\\
$NN\sigma$
&$-g_{NN\sigma} \bar{\Psi} \Psi \sigma $\\
$\sigma\pi\pi$
&$\frac{g_{\sigma \pi \pi}}{2 m_{\pi}} \partial_{\mu} \vec{\pi} \partial^{\mu}
\vec{\pi} \sigma $\\
$\sigma\sigma\sigma$
&$-g_{\sigma \sigma \sigma} m_{\sigma} \sigma \sigma \sigma $\\
$NN\rho\pi$
&$\frac{f_{NN\pi}}{m_{\pi}}g_{\rho}
\bar{\Psi}\gamma^5\gamma^{\mu}\vec{\tau}\Psi(\vec{\rho}_{\mu} \times \vec{\pi})
$\\
$NNa_1$
&$-\frac{f_{NN\pi}}{m_{\pi}}m_{a_1}\bar{\Psi}\gamma^5\gamma^{\mu}\vec{\tau}\Psi
\vec{a}_{\mu} $\\
$a_1\pi\rho$
&$-\frac{g_{\rho}}{m_{a_1}}\left[\partial_{\mu}\vec{\pi}\times\vec{a}_{\nu}-\partial_{\nu}\vec{\pi}\times\vec{a}_{\mu}\right]\left[\partial^{\mu}\vec{\rho}^{\nu}-\partial^{\nu}\vec{\rho}^{\mu}\right]
$\\
&$+\frac{g_{\rho}}{2m_{a_1}}\left[\vec{\pi}\times(\partial_{\mu}\vec{\rho}_{\nu}-\partial_{\nu}\vec{\rho}_{\mu})\right]\left[\partial^{\mu}\vec{a}^{\nu}-\partial^{\nu}\vec{a}^{\mu}\right]
$\\
$NN\omega$
&$-g_{NN\omega}
\bar{\Psi}[\gamma^{\mu}-\frac{\kappa_{\omega}}{2m_N}\sigma^{\mu\nu}\partial_{\nu}]\omega_{\mu}
\Psi $\\
$\omega\pi\rho$
&$\frac{g_{\omega\pi\rho}}{m_{\omega}}\epsilon_{\mu\alpha\lambda\nu}\partial^{\alpha}\vec{\rho}^{\mu}\partial^{\lambda}\vec{\pi}\omega^{\nu}
$\\
$N \Delta\rho$
&$-i\frac{f_{N\Delta\rho}}{m_{\rho}}
\bar{\Delta}^{\mu}\gamma^5\gamma^{\nu}\vec{T}^{\dagger}\vec{\rho}_{\mu\nu}\Psi
+\text{h.c.}$\\
$\rho\rho\rho$
&$\frac{g_{\rho}}{2}(\vec{\rho}_{\mu}\times\vec{\rho}_{\nu})\vec{\rho}^{\mu\nu}
$\\
$NN \rho\rho$
&$\frac{\kappa_{\rho}g^2_{\rho}}{8m_N}\bar{\Psi}\sigma^{\mu\nu}\vec{\tau}\Psi(\vec{\rho}_{\mu}\times\vec{\rho}_{\nu})$
\\
$\Delta\Delta \pi$
&$\frac{f_{\Delta\Delta\pi}}{m_{\pi}}\bar{\Delta}_{\mu} \gamma^5\gamma^{\nu}
\vec{T}\Delta^{\mu}\partial_{\nu}\vec{\pi}$ \\
$\Delta\Delta \rho$
&$-g_{\Delta\Delta\rho}\bar{\Delta}_{\tau}
\left(\gamma^{\mu}-i\frac{\kappa_{\Delta\Delta\rho}}{2m_{\Delta}}\sigma^{\mu\nu}\partial_{\nu}\right)\vec{\rho}_{\mu}
\vec{T}\Delta^{\tau}$ \\
$N^*(S_{11})N\pi$
&$ig_{N^*N\pi} \bar{\Psi}_{N^*}\vec{\tau} \Psi \vec{\pi} + \text{h.c.}$ \\
$N^*(S_{11})N\eta$
&$g_{N^*N\eta} \bar{\Psi}_{N^*} \Psi \eta + \text{h.c.}$ \\
$N^*(S_{11})N\rho$
&$g_{N^*N\rho} \bar{\Psi}_{N^*} \gamma^5
[\gamma^{\mu}-\frac{\kappa_{N^*N\rho}}{2m_{N^*}}\sigma^{\mu\nu}\partial_{\nu}]\vec{\tau}\vec{\rho}_{\mu}
\Psi + \text{h.c.}$ \\
$NN\eta$
&$-\frac{f_{NN\eta}}{m_{\pi}} \bar{\Psi}\gamma^5\gamma^{\mu}
\vec{\tau}\partial_{\mu}\vec{\pi} \Psi $ \\
$NNa_0$
&$g_{NNa_0} m_{\pi} \bar{\Psi} \vec{\tau} \Psi \vec{a}_0$ \\
$NNf_0$
&$g_{NNf_0} m_{\pi} \bar{\Psi} \vec{\tau} \Psi \vec{a}_0$ \\
$\pi\eta a_0$
&$g_{\pi\eta a_0} m_{\pi} \eta \vec{\pi} \vec{a}_0$ \\
$\eta\eta f_0$
&$g_{\eta\eta f_0} m_{\pi} \eta \eta f_0$ \\
$N^*(D_{13})N\pi$
&$i\frac{f_{N^*N\pi}}{m^2_{\pi}} \bar{\Psi} \gamma^5 \gamma^{\nu} \vec{\tau}
\Psi_{N^*}^{\mu}\partial_{\nu}\partial_{\mu} \vec{\pi} + \text{h.c.}$ \\
$N^*(D_{13})N\eta$
&$i\frac{f_{N^*N\eta}}{m^2_{\pi}} \bar{\Psi} \gamma^5 \gamma^{\nu}
\Psi_{N^*}^{\mu}\partial_{\nu}\partial_{\mu} \eta + \text{h.c.}$ \\
$N^*(D_{13})\Delta\pi$
&$\frac{f_{N^*\Delta\pi}}{m_{\pi}} \bar{\Psi}_{N^*\nu} \vec{T} \gamma^{\mu}
\Delta^{\nu} \partial_{\mu} \vec{\pi} +\text{h.c.}$ \\
$N^*(D_{13})N\rho$
&$-i\frac{f_{N^*N\rho}}{m_{\rho}}
\bar{\Psi}_{N^*}^{\mu}\gamma^{\nu}\vec{\tau}\vec{\rho}_{\mu\nu}\Psi
+\text{h.c.}$\\
\end{tabular}
\caption{The effective Lagrangian}\label{tablag}
\end{center}
\end{table}

\begin{table}
\begin{center}
\begin{tabular}{|l l l c c|}
vertex & process & coupling const. & Ref. & cutoff $\Lambda$ \\
\hline
correlated $2\pi$-- & $\rho$--channel & && {\bf 1200} \\
exchange     & $\sigma$--channel & && {\bf 1100} \\
\hline
$NN\pi$ & $N$ exchange & $\frac{f^2_{NN\pi}}{4\pi}=0.0778$ &\cite{Janssen96}&
{\bf 1300} \\
$NN\pi$ & $N$ pole, $m^0_N=1032.33$ & $\frac{f^{(0)\;2}_{NN\pi}}{4\pi}=0.0633$
&& {\bf 1200} \\
$N\Delta\pi$ & $N$ exchange & $\frac{f^2_{N\Delta\pi}}{ 4\pi}=0.36$
&\cite{Janssen96}& {\bf 1300} \\
$N\Delta\pi$ & $\Delta$ exchange & $\frac{f^2_{N\Delta\pi}}{ 4\pi}=0.36$ &
\cite{Janssen96}& {\bf 1800} \\
$N\Delta\pi$ & $\Delta$ pole, $m^0_\Delta ={\bf 1405}$ &
$\frac{f^{(0)\;2}_{N\Delta\pi}}{ 4\pi}={\bf 0.21}$ && {\bf 1650} \\
$\Delta\Delta\pi$ & $\Delta$ exchange & $\frac{f^2_{\Delta\Delta\pi}}{
4\pi}=0.252$ & \cite{Schuetz95J,Brown75}& {\bf 1800} \\
$N\Delta\rho$ & $\rho$ exchange & $\frac{f^2_{N\Delta\rho}}{ 4\pi}=20.45$ &
\cite{Janssen96}& {\bf 1300} \\
$\Delta\Delta\rho$ & $\rho$ exchange & $\frac{g^{V\,2}_{\Delta\Delta\rho}}{
4\pi}=4.69$, &\cite{Schuetz95J,Brown75}& {\bf 1300} \\
 & & $\frac{g^T_{\Delta\Delta\rho} }{ g^V_{\Delta\Delta\rho}}=6.1$ &
\cite{Schuetz95J,Brown75}& \\
$\pi\pi\rho$ & $\rho$ exchange & $\frac{g^2_{\rho\pi\pi}}{ 4\pi}=2.90$ &
\cite{Janssen95}& {\bf 1300} \\
\hline
$NN\sigma$ & $N$ exchange & $\frac{g^2_{NN\sigma}}{ 4\pi}=13$ &\cite{Durso80}&
{\bf 1500} \\
$NN\pi$ & $\pi$ exchange &  $\sim{f_{NN\pi}}$ && {\bf 600} \\
$\pi\pi\sigma$ & $\pi$ exchange & $\frac{g^2_{\pi\pi\sigma}}{ 4\pi}=0.25$ &
\cite{Krehl97}& {600} \\
$NN\sigma$ & $\sigma$ exchange & $\sim{g_{NN\sigma}}$ && {\bf 2300} \\
$\sigma\sigma\sigma$ & $\sigma$ exchange & $\frac{g^2_{\sigma\sigma\sigma}}{
4\pi}={\bf 0.625}$ && {2300} \\
\hline
$NN\eta$ & $N$ exchange & $\frac{f^2_{NN\eta}}{ 4\pi}=0.00934$ &
\cite{Schuetz98}& {\bf 2500} \\
$NNa_0$ & $a_0$ exchange & $ \frac{g_{NNa_0}g_{\pi\eta a_0}}{ 4\pi}={\bf 8.0}$
&& {\bf 2500} \\
$\pi\eta a_0$ & $a_0$ exchange &  &&  2500 \\
\hline
$NN\rho$ & $N$ exchange & $\frac{g^2_{NN\rho}}{ 4\pi} =0.84$ &
\cite{Janssen96}& {\bf 1200} \\
 &  & $\kappa=6.1$ & \cite{Janssen96}&  \\
$NN\rho\pi$ & contact term & $\sim f_{NN\pi}g_{NN\rho}$ && {\bf 1100} \\
$\pi\pi\rho$ & $\pi$ exchange & $\frac{g^2_{\pi\pi\rho}}{ 4\pi}$ && {600} \\
$NN\omega$ & $\omega$ exchange & $\frac{g^2_{NN\omega}}{ 4\pi}=11.0$ &
\cite{Janssen96}& {\bf 1100} \\
$\omega\pi\rho$ & $\omega$ exchange & $\frac{g^2_{\omega\pi\rho}}{ 4\pi}=10.0$
&\cite{Nakayama98,Durso87}& {\bf 700} \\
$NNa_1$ & $a_1$ exchange & $\sim f_{NN\pi}$&& {\bf 1500} \\
$a_1\pi\rho$ & $a_1$ exchange & $\sim g_{NN\rho}$&& {1500} \\
$NN\rho$ & $\rho$ exchange & $g_{NN\rho},\kappa$ && {\bf 1400} \\
$\rho\rho\rho$ & $\rho$ exchange &  $\sim g_{NN\rho}$ && {1400} \\
$NN\rho\rho$ & contact term & $\sim g^2_{NN\rho}\kappa$ && {\bf 1200} \\
\hline
$NN^{*\,S_{11}}_{1535}\pi$ & $N^*$ pole, $m^0_{N^*}={\bf 1660}$ &
$\frac{g^2_{NN^*\pi}}{ 4\pi} ={\bf 0.0015}$ && {\bf 3000} \\
$NN^{*\,S_{11}}_{1535}\eta$ & $N^*$ pole  & $\frac{g^2_{NN^*\eta}}{ 4\pi} ={\bf
0.30}$ && {3000} \\
$NN^{*\,S_{11}}_{1650}\pi$ & $N^*$ pole, $m^0_{N^*}={\bf 1852}$ &
$\frac{g^2_{NN^*\pi}}{ 4\pi} ={\bf 0.08}$ && {\bf 3000} \\
$NN^{*\,S_{11}}_{1650}\rho$ & $N^*$ pole  & $\frac{g^2_{NN^*\rho}}{ 4\pi} ={\bf
0.05}$ && {3000} \\
$NN^{*\,D_{13}}_{1520}\pi$ & $N^*$ pole, $m^0_{N^*}={\bf 2100}$ &
$\frac{f^2_{NN^*\pi}}{ 4\pi} ={\bf 0.0006}$ && {\bf 2000} \\
$NN^{*\,D_{13}}_{1520}\rho$ & $N^*$ pole  & $\frac{f^2_{NN^*\rho}}{ 4\pi} ={\bf
0.20}$ && {2000} \\
$\Delta N^{*\,D_{13}}_{1520}\pi$ & $N^*$ pole,  & $\frac{f^2_{\Delta N^*\pi}}{
4\pi} ={\bf 0.017}$ && {2000} \\
$N N^{*\,D_{13}}_{1520}\eta$ & $N^*$ pole  & $\frac{f^2_{N N^*\eta}}{ 4\pi}
={\bf 0.0008}$ && {2000}

\end{tabular}
\caption{The parameters of our model. Only the boldface printed values are
varied in fitting
  the data. The coupling constants are taken from the cited
  references. All masses and cutoffs are given in MeV.}
\label{tabparams}
\end{center}
\end{table}

\begin{table}
\begin{center}
\begin{tabular}{|ll|ll|ll|}
mesons && baryons && exchanged mesons &\\
\hline
$m_{\pi}$ & 138.03 &$m_{N}$ & 938.926 &$m_{\sigma}$ & 650.0$^a$ \\
$m_{\eta}$ & 547.45 &$m_{\Delta}$ & 1232.0 &$m_{\omega}$ & 782.6 \\
$m_{\sigma}$ & 850.0$^a$ & & 1520.0 &$m_{f_0}$ & 974.1 \\
$m_{\rho}$ & 769.0 & & &$m_{a_0}$ & 982.7 \\
&&&&$m_{a_1}$ & 1260.0 \\
\end{tabular}
\caption{Masses of the mesons and baryons (in MeV).\quad$^a$ The $\sigma$ mass
in the $s$-channel
$\pi\pi$ interaction corresponds to the energy at which the phase shift reaches
90$^o$.
The $\sigma$ in the $\sigma N$ $t$-channel exchange is a parameterization of
correlated $\pi\pi$
exchange \protect \cite{Durso80}. This is the reason for the different $\sigma$
masses.}
\label{tabmassen}
\end{center}
\end{table}

\begin{table}
\begin{center}
\begin{tabular}{|lllll|}
set & $\frac{f^2_{N\pi}}{4\pi}$ & $\frac{f^2_{N\sigma}}{4\pi}$&
$\frac{f^2_{\Delta\pi}}{4\pi}$ & $m^0$ (in MeV)\\
\hline
I & 0.024 & 20.21 & 0 & 2840 \\
II & 0.024 & 0 & 0.17 & 3950 \\
III & 0.018 & 0 & 0.20 & 4100 
\end{tabular}
\caption{Parameters of the separable coupled-channel model}
\label{sepparams}
\end{center}
\end{table}

\begin{table}
\begin{center}
\begin{tabular}{|c|c|c|c|}
reaction channel & process & $IF(I=1/2)$&$IF(I=3/2)$ \\
\hline
$\pi N \to \pi N$ & $\sigma$ exchange & $1$ & $1$  \\
                  & $\rho$ exchange & $2$ & $-1$  \\
                  & $N^*_{D_{13}}$ pole graph & $3$ & $0$  \\
$\pi N \to \rho N$ & $N$ exchange & $-1$ & $2$  \\
                   & $NN\pi\rho$ contact graph & $-2i$ & $i$  \\
                   & $\pi$ exchange & $-2i$ & $i$  \\
                   & $\omega$ exchange & $1$ & $1$  \\
                   & $a_1$ exchange & $-2i$ & $i$  \\
                   & $\Delta$ exchange & $\frac{4}{3}$ & $\frac{1}{3}$  \\
                   & $N^*_{S_{11}},N^*_{D_{13}}$ pole diagrams & $3$ & $0$  \\
$\rho N \to \rho N$& $N$ exchange & $-1$ & $2$  \\
                   & $NN\rho\rho$ contact graph & $-2i$ & $i$  \\
                   & $\rho$ exchange & $2i$ & $-i$  \\
                   & $\Delta$ exchange & $\frac{4}{3}$ & $\frac{1}{3}$  \\
                   & $N^*_{S_{11}},N^*_{D_{13}}$ pole diagrams & $3$ & $0$  \\
$\pi N \to \sigma N$ & $\pi$ exchange & $\sqrt{3}$ & $0$  \\
$\pi N \to \pi\Delta$ & $N^*_{D_{13}}$ pole diagram & $-\sqrt{6}$ & $0$  \\
$\pi \Delta \to \pi\Delta$ & $N^*_{D_{13}}$ pole diagram & $2$ & $0$  \\
$\pi N \to \eta N$ & $N^*_{D_{13}}$ pole graph & $\sqrt{3}$ & $0$  \\
$\eta N \to \eta N$ & $N^*_{D_{13}}$ pole graph & $1$ & $0$  \\
\end{tabular}
\caption{Additional isospin factors}
\label{tabiso}
\end{center}
\end{table}

\newpage

\begin{figure}[h]
\begin{center}
\epsfig{file=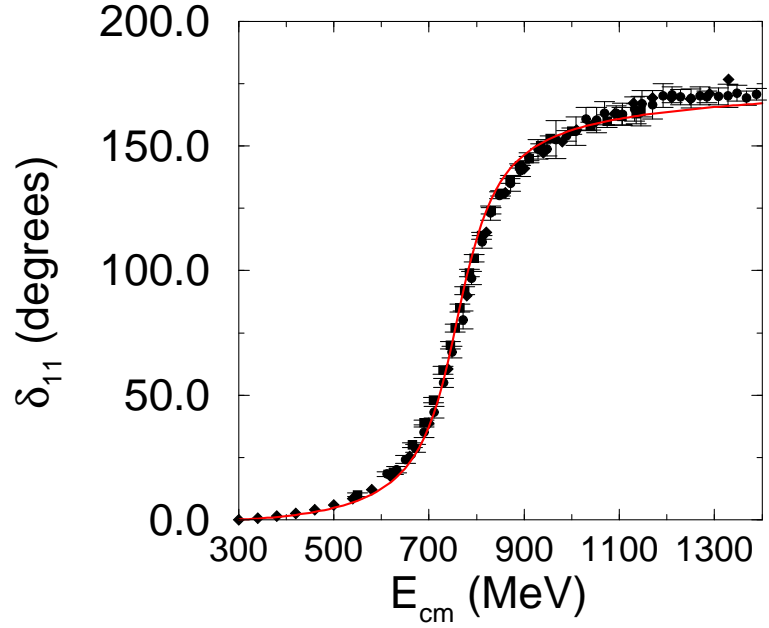,width=10cm}
\caption{Phase shift in the partial wave $IJ=11$ of the $\pi\pi$ interaction.
The solid line is the result of the self-energy calculation for the $\rho$
meson.  Data is taken from Refs. \protect \cite{Ochs73,Froggatt77,Hyams75}.}
\label{selfrho}
\end{center}
\end{figure}

\begin{references}


\bibitem{Koch85}{R. Koch},
Z. Phys. {\bf {C 29}},  597  (1985).
\bibitem{SM95} {R.A. Arndt, I.I. Strakovsky, R.L. Workman, and M.M. Pavan},
Phys. Rev. C {\bf 52},  2120  (1995).
\bibitem{SP98}{R.A. Arndt, R.L. Workman, I.I. Strakovsky, and M.M. Pavan},
{LANL preprint nucl-th/9807087}  (1998).
\bibitem{Batinic95} {M. Batini\`c, I. $\check{S}$laus, A. $\check{S}$varc, and
B.M.K. Nefkens},
Phys. Rev. C {\bf 51}, 2310 (1995).
\bibitem{Feuster99} {T. Feuster and U. Mosel},
Phys. Rev. C {\bf 59},  460  (1999).
\bibitem{Vrana99}{T.P. Vrana, S.A. Dytman, and T.-S.H. Lee},
LANL preprint nucl-th/9910012.
\bibitem{Cutkosky79}{R.E. Cutkosky et al.},
Phys. Rev. D {\bf 20},  {2804 and 2839}  (1979).
\bibitem{Manley92}{D.M. Manley and E.M. Saleski},
Phys. Rev. D {\bf 45},  4002  (1992).
\bibitem{PDG98} {C. Caso et al.},
Eur. Phys. J. C {\bf 3}, 1 (1998).
\bibitem{Zou99} {B.S. Zou, G.X. Peng, H.C. Chiang, and P.N. Shen},
LANL preprint hep-ph/9909204.
\bibitem{Isgur78} {N. Isgur and G. Karl},
Phys. Rev. D {\bf 18}, 4187 (1978).
\bibitem{Isgur79} {N. Isgur and G. Karl},
Phys. Rev. D {\bf 19}, 2653 (1979).
\bibitem{Capstick86} {S. Capstick and N. Isgur},
Phys. Rev. D {\bf 34}, 2809 (1986).
\bibitem{Glozman96} {L.Ya. Glozman and D.O. Riska},
Phys. Rep. {\bf 268}, 263 (1996).
\bibitem{Bijker97} {R. Bijker, F. Iachello, and A. Leviatan},
Phys. Rev. D {\bf 558}, 2862 (1997).
\bibitem{Liu99} {K.F. Liu, S.J. Dong, T. Draper, D. Leinweber, J. Sloan, W.
Wilox, and R.M. Woloshyn},
Phys. Rev. D {\bf 59}, 112001 (1999).
\bibitem{Lee99}{T. Yoshimoto, T. Sato, M. Arima, and T.-S. H. Lee},
LANL preprint nucl-th/9908048.
\bibitem{Isgur99}{N. Isgur},
LANL preprint nucl-th/9908028.
\bibitem{Glozman99}{L.Ya. Glozman},
LANL preprint nucl-th/9909021.
\bibitem{Li90}{Z. Li and F.E. Close},
Phys. Rev. D {\bf 42}, 2207 (1990).
\bibitem{Li92}{Z. Li, V. Burkert, and Z. Li},
Phys. Rev. D {\bf 46}, 70 (1992).
\bibitem{Capstick92}{S. Capstick},
Phys. Rev. D {\bf 46}, 1965 (1992).
\bibitem{Capstick95}{S. Capstick and B.D. Keister},
Phys. Rev. D {\bf 51}, 3598 (1995).
\bibitem{Cardarelli97}{F. Cardarelli, E. Pace, G. Salmgrave{e}, and S. Simula},
Phys. Lett. {\bf B397}, 13 (1997).
\bibitem{Dong99}{Y.B. Dong, K. Shimizu, A. Faessler, and A.J. Buchmann},
Phys. Rev. C {\bf 60}, 035203 (1999).
\bibitem{Capstick93} {S. Capstick and W. Roberts},
Phys. Rev. D {\bf 47}, 1994 (1993).
\bibitem{Capstick94} {S. Capstick and W. Roberts},
Phys. Rev. D {\bf 49}, 4570 (1994).
\bibitem{Stancu89} {Fl. Stancu and P. Stassart},
Phys. Rev. D {\bf 41}, 916 (1990).
\bibitem{Stancu90} {Fl. Stancu and P. Stassart},
Phys. Rev. D {\bf 39}, 343 (1989).
\bibitem{Stassart92} {P. Stassart},
Phys. Rev. D {\bf 46}, 2085 (1992).
\bibitem{Cutkosky90}{R.E. Cutkosky and S. Wang},
Phys. Rev. D {\bf 42},  235  (1990).
\bibitem{Barnes94} {T. Barnes},
RAL-94-056 and LANL preprint hep-ph/9406215.
\bibitem{Maltman86} {K. Maltman and S. Godfrey},
Nucl. Phys. {\bf A452} 669 (1986).
\bibitem{Stancu98} {Fl. Stancu},
Phys. Rev. C {\bf 58} 111501 (1998).
\bibitem{Genovese98} {M. Genovese, J.-M. Richard, Fl. Stancu, and S. Pepin},
Phys. Lett. {\bf B425} 171 (1998).
\bibitem{Hoehler92}{G. H\"ohler and A. Schulte},
$\pi N$ Newsletter {\bf 7},  94  (1992).
\bibitem{SM90}{R.A. Arndt, Z. Li, L.D. Roper, R.L. Workman, and J.M. Ford},
Phys. Rev. D {\bf 43},  2131  (1991).
\bibitem{Hoehler98} G. H\"ohler, {\it Note on $N$ and $\Delta$ resonances II.
Against Breit-Wigner parameters -- a pole-emic} in Ref. \cite{PDG98}.
\bibitem{KA84}
{G. H\"ohler}, {private communication}.
\bibitem{Blankleider85}{B. Blankleider and G.E. Walker},
Phys. Lett.{\bf B152}, 291 (1985).
\bibitem{Fuda98}{Y. Elmessiri and M.G. Fuda},
Phys. Rev. C {\bf 57} 2149 (1998).
\bibitem{Gridnev99} {A.B. Gridnev and N.G. Kozlenko},
Eur. Phys. J. A {\bf 4},  187  (1999).
\bibitem{Joachain75} {C.J. Joachain}, {\it Quantum Collision Theory}
(North-Holland Publihing Company, Amsterdam, 1975).
\bibitem{Machleidt87} R. Machleidt, K. Holinde, and Ch. Elster,
Phys. Rep. 149, 1 (1987).
\bibitem{Badalyan82} {A.M. Badalyan et al.},
Phys. Rep. {\bf 82}, 31 (1982).
\bibitem{Weinstein90} {J. Weinstein and N. Isgur},
Phys. Rev. D {\bf 41}, 2236 (1990).
\bibitem{Janssen95} {G. Jan\ss{}en, B.C. Pearce, K. Holinde, and J. Speth},
Phys. Rev. D {\bf 52}, 2690 (1995).
\bibitem{Siegel88} {P.B. Siegel and W. Weise},
Phys. Rev. C {\bf 38}, 2221 (1988).
\bibitem{Groeling90}{A.~M\"uller-Groeling, K.~Holinde, and J.~Speth},
Nucl.\ Phys.\ {\bf A513}, 557 (1990).
\bibitem{Schuetz98}
{C. Sch\"utz, J. Haidenbauer, J.W. Durso, and J. Speth}, Phys. Rev. C {\bf 57},
  1464  (1998).
\bibitem{Bockmann99} {R. B\"ockmann, C. Hanhart, O. Krehl, S. Krewald, and J.
Speth},
Phys. Rev. C {\bf 60},  055212  (1999).
\bibitem{Pearce91}
{B.C. Pearce and B.K. Jennings}, Nucl. Phys. {\bf A528},  655  (1991).
\bibitem{Lee91}
{C. Lee, S.N. Yang, and T.-S. H. Lee}, J. Phys. G {\bf 17},    (1991).
\bibitem{Gross93}
{F. Gross and Y. Surya}, Phys. Rev. C {\bf 47},  703  (1993).
\bibitem{Hung94}
{C.T. Hung, S.N. Yang, and T.-S. H. Lee}, {J. Phys. G} {\bf 20},  1531  (1994).
\bibitem{Goudsmit94} {P.F.A. Goudsmit, H.J. Leisi, E. Matsinos, B.L. Birbrair,
and A.B. Gridnev},
Nucl. Phys. {\bf A575},  673  (1994).
\bibitem{Pascalutsa98}
{V. Pascalutsa and J.A. Tjon}, Nucl. Phys. {\bf A631},  534c  (1998).
\bibitem{Lahiff99} {A.D.~Lahiff and I.R.~Afnan},
LANL preprint nucl-th/9903058.
\bibitem{Buettgen90}{R.~Buettgen, K.~Holinde, A.~Mueller-Groeling, J.~Speth and
P.~Wyborny},
Nucl.\ Phys.\ {\bf A506}, 586 (1990).
\bibitem{Wess67}
{J. Wess and B. Zumino}, Phys. Rev. {\bf 163},  1727  (1967).
  2671  (1994).
\bibitem{Krehl99} {O. Krehl, C. Hanhart, S. Krewald, and J. Speth},
Phys. Rev. C {\bf 60},  055206  (1999).
\bibitem{Schuetz95}{C. Sch\"utz, K. Holinde, J. Speth, B.C. Pearce und J.W.
Durso},
Phys. Rev. C {\bf 51},  1374  (1995).
\bibitem{Erkelenz74}{K. Erkelenz},
Phys. Rep. {\bf 5}, 191 (1974).
\bibitem{Hetherington65}
{J.H. Hetherington and L.H. Schick}, Phys. Rev. B {\bf 137},  935  (1965).
\bibitem{Aaron66}
{R. Aaron and R.D. Amado}, Phys. Rev. {\bf 150},  857  (1966).
\bibitem{Cahill71}
{R.T. Cahill and I.H. Sloan}, Nucl. Phys. {\bf A165},  161  (1971).
\bibitem{JacobWick59}
{M. Jacob and G.C. Wick}, Ann. Phys. (N.Y.) {\bf 7},  404  (1959).
\bibitem{Schuetz94}
{C. Sch\"utz, J.W. Durso, K. Holinde, and J. Speth}, Phys. Rev. C {\bf 49},
\bibitem{Hoehler83}
{G. H\"ohler}, {in {\it Pion-Nucleon Scattering}, edited by H. Schopper,
  Landolt-B\"ornstein, New Series, Group I, Vol. 9b, Pt. 2 (Springer-Verlag,
  New York, 1983)}.
\bibitem{Afnan98}{I.R. Afnan},
Acta Phys. Polon. B {\bf 29}, 2397 (1998).
\bibitem{Elsey89}{J.A. Elsey and I.R. Afnan},
Phys. Rev. D {\bf 40}, 2353 (1989).
\bibitem{Ochs73}
{W. Ochs}, {Thesis, unpublished (M\"unchen, 1973)}.
\bibitem{Froggatt77}
{C.D. Froggatt and J.L. Petersen}, Nucl. Phys. {\bf B129},  89  (1977).
\bibitem{Hyams75}
{B. Hyams et al.}, Nucl. Phys. {\bf B100},  205  (1975).
\bibitem{Janssen96}
{G. Jan\ss{}en, K. Holinde, and J. Speth}, Phys. Rev. C {\bf 54},  2218
 (1996).
\bibitem{Schuetz95J}
{C. Sch\"utz}, {Berichte des Forschungszentrums J\"ulich, Nr. 3130 (1995)}.
\bibitem{Brown75}
{G.E. Brown and W. Weise}, Phys. Rep. {\bf 22},  279  (1975).
\bibitem{Durso80}
{J.W. Durso, A.D. Jackson and B.J. Verwest}, Nucl. Phys. {\bf {A345}},  471
  (1980).
\bibitem{Krehl97}
{O. Krehl and J. Speth}, Nucl. Phys. {\bf A623},  162c  (1997).
\bibitem{Nakayama98}
{K. Nakayama, A. Szczurek, C. Hanhart, J. Haidenbauer, and J. Speth}, Phys.
Rev.
  C {\bf 57},  1580  (1998).
\bibitem{Durso87}
{J.W. Durso}, Phys. Lett. B {\bf 184},  348  (1987).
\bibitem{LB87}
{\em Handbook of Physics, Landolt--B\"ornstein}, edited by H. Schopper
(Springer,
Berlin, 1987), Vol.~12/a.
\bibitem{Aaron68}
{R. Aaron, R.D. Amado und J.E. Young.}, Phys. Rev. {\bf 174},  2022  (1968).
\bibitem{Aaron69}
{R. Aaron, D.C. Teplitz, R.D. Amado und J.E. Young}, Phys. Rev. {\bf 187},
  2047  (1969).
\bibitem{Krehl2000}{O. Krehl, C. Hanhart, S. Krewald, and J. Speth},
in preparation.
\bibitem{Goldberger64}
{M.L. Goldberger and K.M. Watson}, {\em {Collision Theory}} ({John Wiley and
  Sons}, {New York}, 1964).
\bibitem{Bohm93}
{A. Bohm}, {\em {Quantum Mechanics: Foundations and Applications}}
  ({Springer--Verlag}, {New York}, 1993).
\bibitem{Hoehler93}
{G. H\"ohler}, $\pi$N Newsletter {\bf 9},  1  (1993).
\bibitem{McLeod85}
{R.J. McLeod and I.R. Afnan}, Phys. Rev. C {\bf 32},  222  (1985).
\bibitem{Afnan81}
{I.R. Afnan and A.T. Stelbovics}, Phys. Rev. C {\bf 23},  1384  (1981).
\bibitem{Frazer60} W.R. Frazer and J.R. Fulco,
Phys. Rev. {\bf 117}, 1609 (1960).
\bibitem{GrossBook}
{F. Gross}, {\em Relativistic Quantum Mechanics and Field Theory}
  ({John Wiley \& Sons}, {New York}, 1993).
\bibitem{Schweber62}
{S.S. Schweber}, {\em {An Introduction to relativistic Quantum Field Theory}}
  ({Harper and Row}, {New York}, 1962).


\end{references}
\end{document}